\documentclass[aps,showpacs,epsf,two column,pra]{revtex4}
\usepackage{amssymb}
\usepackage{amsmath}
\usepackage{graphicx,psfrag}
\usepackage{subfig}
\usepackage{float}
\usepackage{tikz}
\usepackage{pgfplots}
\usepackage{braket}
\usepackage{soul}
\usepackage[colorlinks=true, citecolor=blue, urlcolor = blue, linkcolor= red, bookmarks=true]{hyperref}
\makeatletter
\renewcommand*\env@matrix[1][*\c@MaxMatrixCols c]{%
  \hskip -\arraycolsep
  \let\@ifnextchar\new@ifnextchar
  \array{#1}}
\makeatother

\makeatletter
   \def\vhrulefill#1{\leavevmode\leaders\hrule\@height#1\hfill \kern\z@}
\makeatother
\usepackage{mleftright}

\begin{document}
\def \beq{\begin{equation}}
\def \eeq{\end{equation}}
\def \bse{\begin{subequations}}
\def \ese{\end{subequations}}
\def \bea{\begin{eqnarray}}
\def \eea{\end{eqnarray}}
\def \bem{\begin{displaymath}}
\def \eem{\end{displaymath}}
\def \bem{\begin{bmatrix}}
\def \eem{\end{bmatrix}}
\def \Ps{\hat{\Psi}(\boldsymbol{r})}
\def \Pds{\hat{\Psi}^{\dagger}(\boldsymbol{r})}
\def \i{{\int}d^2{\bf r}}
\def \bl{\bar{\boldsymbol{l}}}
\def \c{\hat{c}_{n,m}}
\def \cp{\hat{c}_{n',m'}}
\def \cd{\hat{c}_{n,m}^{\dagger}}
\def \cdp{\hat{c}_{n',m'}^{\dagger}}
\def \bb{\bibitem}
\def \nn{\nonumber}

\def \bs{\boldsymbol}
\def \hkx{\hat{k}_{x}}
\def \hky{\hat{k}_{y}}
\def \bq{\bar{q_{y}}}

\def \bc{\begin{center}}
\def \ec{\end{center}}

\title{\textbf{ $(2+1)$-dimensional sonic black hole from spin-orbit coupled  Bose-Einstein condensate and its analogue Hawking radiation}}
\author{Inderpreet Kaur and Sankalpa Ghosh}
\affiliation{Department of Physics, Indian Institute of Technology Delhi, New Delhi-110016, India}

\begin{abstract}
We study the properties of a $2+1$ dimensional Sonic black hole (SBH) that can be realised, in a quasi-two-dimensional two-component spin-orbit coupled Bose-Einstein condensate (BEC). The corresponding equation for phase fluctuations in the total density mode that describes phonon field in the hydrodynamic approximation is described by a scalar field equation in $2+1$ dimension whose space-time metric is significantly different from that of the SBH realised from a single component BEC that was studied experimentally, and, theoretically meticulously in literature. Given the breakdown of the irrotationality constraint of the velocity field in such spin-orbit coupled BEC, we study in detail how the time evolution of such condensate impacts the various properties of the resulting SBH. 
By time evolving the condensate in a suitably created laser-induced potential, we show that such a sonic black hole is formed, in an annular region bounded by inner and outer event horizon as well as elliptical ergo- surfaces. We observe amplifying density modulation due to the formation of such sonic horizons and show how they change the nature of analogue Hawking radiation emitted from such sonic black hole by evaluating the density-density correlation at different times, using the truncated Wigner approximation (TWA) for different values of spin-orbit coupling parameters. We finally investigate the thermal nature of such analogue Hawking radiation. 

\end{abstract}

\maketitle

\newpage

\section{Introduction}
The  Hawking radiation \cite{ Hawking1, Hawking2} is predicted to be emitted from a black hole (BH), 
formed at a particular stage of stellar evolution, and combines the principle of General Theory of Relativity (GTR) with those of Quantum field theory. However, direct measurement in real BHs to observe such radiation is less likely even in the foreseeable future. Analogue systems in quantum fluids that can kinematically simulate Hawking radiation \cite{Unruh}, therefore, attract a lot of research interest.
Recent observation of analogue Hawking radiation (HR) from a sonic black hole (SBH)
in ultra-cold atomic superfluid of a Bose-Einstein condensate (BEC) of $^{87}$Rb \cite{Steinhauer14, Steinhauer16} is a major step towards this direction. 
It was followed by another experiment where a sonic analogue of expanding universe was realised in a supersonically expanding ring-shaped $^{23}$Na BEC \cite{Eckel}. Ultracold atomic systems thus emerged as a frontier candidate to test phenomena related to Gravitation and Cosmology through analogue experiments.

Analogue SBH in ultra-cold superfluid exists due to the fact 
that the hydrodynamic equation of phonons, which are the low energy quasiparticles of such atomic superfluid, 
takes a covariant form with a curved space-time metric, mimicking the curved space-time in GTR near a BH \cite{Unruh, visser90} given by: 
\begin{equation}
\frac{1}{\sqrt{-g}}\partial_{\mu}( \sqrt{-g} g^{\mu\nu}\partial_{\nu} \tilde{\theta})=0 \label{phononeqs}
 \end{equation}
where $\tilde{\theta}$ is the fluctuation in the phase of the superfluid BEC and $g^{\mu \nu}$  is the analogue space-time metric. 
The covariant space-time metric $g_{\mu\nu}$ in ``d" spatial dimensions is given as \cite{visser90}:
\begin{equation}
\renewcommand\arraystretch{1.3}
g_{\mu\nu} = \bigg(\frac{n}{c}\bigg)^{\frac{2}{d-1}} \mleft[
\begin{array}{c|c}
   - ({c}^2 -|{\bs v}|^2) &  -{\bs v}^T   \\ 
  \hline
  -{\bs v} &  {\bf I}_{d \times d}
\end{array}
\mright]
\label{a}
\end{equation}
where, c is the sound speed, n is the density and ${\bs v}$ is the flow velocity of the condensate.

 Recent observation of the analogue Hawking radiation \cite{Steinhauer16, Steinhauer18} in an effective $1+1$ dimensional quasi-condensate is a very significant step in understanding the validity of Hawking's treatment to derive the thermal radiation from the black hole.
The background induced metric $g^{\mu \nu}$ 
of such SBH realised in recent experiments \cite{Steinhauer16, Steinhauer18} is described by a $1+1$ dimensional generalisation of static singular Schwarzschild metric \cite{SSch}, which describes a simple curved space-time in GTR.  
However, the strong spatial confinement needed to realise quasi-one-dimensional condensates in such experiments may limit the hydrodynamic modes to reduced degrees of freedom in lower dimensions and the space-time analogy discussed in the preceding paragraph, is associated with such modes only. On the other hand, the proposed sonic analogy for curved space-time, by Unruh \cite{Unruh}, depends on the role of \emph{spatial} dimensions. A more rigorous validation of such sonic analogy thus requires a minimum of two spatial dimensions, as the form of $g^{\mu \nu}$ [Eq.(\ref{a})] in $1+1$ dimensions is ill-defined, since the \textit{conformal pre-factor} in d-spatial dimensions $\sim$ $(n/c)^{\frac{2}{d-1}}$ is singular in $1+1$ dimensions \cite{visser11,anagbook,Anglin}.  This motivates one to go beyond $1+1$ dimensional systems and investigate the evolution of the SBH in higher spatial dimensions. 
Such studies are significantly less in number \cite{Anglin}. 

\begin{figure*}
\includegraphics[scale=0.6]{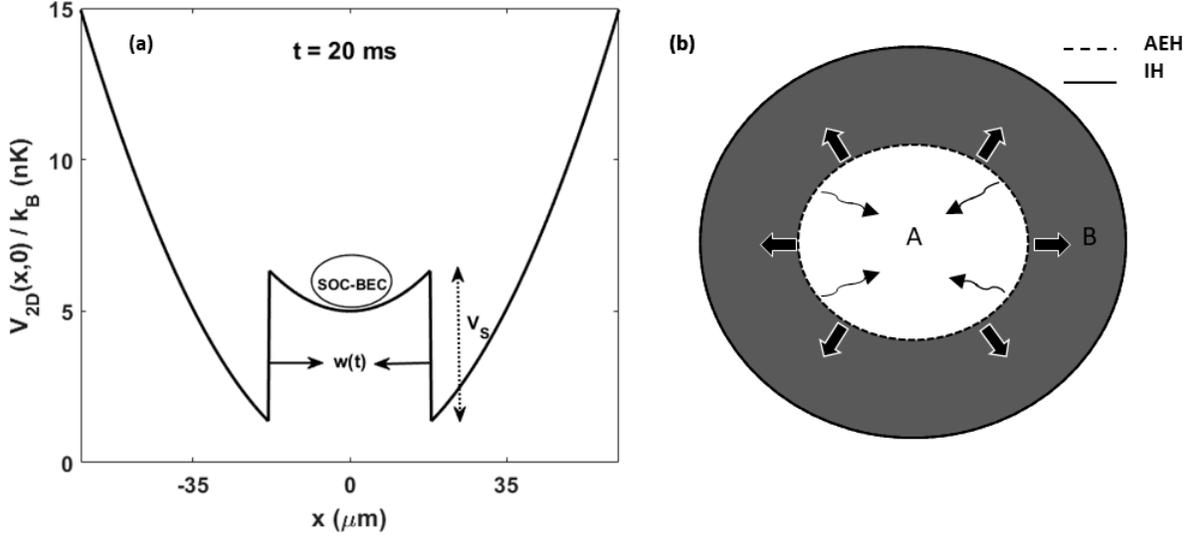}
\centering
\caption{{\it (color online).} (a) Cross-section of the potential, $V_{2D}$ along the x-axis, experienced by the SOC-BEC at 20 ms. (b) Schematic illustration of the formation of the supersonic region (area B) and the subsonic region (area A) of the analogue black hole formed in this work. Region A represents the outside of the analogue black hole and shaded region B represents the inside of the analogue black hole. Thus, the emission of the analogue Hawking radiation from this SBH takes place in region A (indicated using curved arrows). AEH and IH are also marked. The arrows in region B represents that the flow is from AEH to IH inside the SBH.}\label{schematic}
\end{figure*}

In this work, we have considered a sonic black hole creation in the spin-orbit coupled (SOC) Bose-Einstein condensate (BEC)\cite{brandon, spielmannature, Dalibardcoll, socdalibard}, using a suitable time-dependent potential depicted in Fig. \ref{schematic},
to realise a  $2+1$ dimensional geometry for SBH. Apart from providing a natural platform to study the dynamics of a $2+1$ dimensional SBH and the analogue sonic Hawking radiation from such SBH, the SOC-BEC adds another important aspect to the related analogue black hole model which is as follows. Most of the quasi-one-dimensional BEC's that dominate the current study of SBH \cite{Steinhauer14, Steinhauer16, Steinhauer18,entanglementJS,Pavloff20,Zoller,la} in ultracold atomic systems are obtained from a typical three-dimensional BECs under suitable trapping condition that behaves like superfluid with a velocity field $\bs{v}$=$\frac{\hbar}{m} \bs{\nabla }\Phi$ which is irrotational, 
where  $\Phi$ is the phase of the superfluid order parameter \cite{Helium,Lifshitz} and $m$  is the atomic mass. 
Therefore, a considerable azimuthal flow in such superfluid is not achievable unless they are rotated externally at sufficient angular velocities and introduce vorticity \cite{rotatingsoc}, as achieved in several analogue mediums \cite{vortexkerr, vortexhuang, relBEC, photon, superradiant}. Thus, most of the current model of SBH is based, on ultracold condensate whose mean-field (ground state) wave function does not have a finite value of the angular momentum.

However, the aforementioned irrotationality condition in BEC gets violated in the presence of spin-orbit coupling \cite{Vorinsoc, SOexpansion}. For the similar external conditions (e.g., external potential, interaction strength), as compared to a single-component (henceforth called scalar) BEC, the expansion of such SOC-BEC in free space is anisotropic. It is due to the presence of gauge fields  \cite{SOexpansion}, as the effective momentum along a particular direction gets modified depending on the type and strength of spin-orbit coupling resulting in an anisotropic velocity profile. Thus, $g^{\mu \nu}$ in Eq.(\ref{phononeq}) of such SOC-BEC promises a more exotic analogue space-time as compared to the scalar BEC. The subsequent simulation shown in this work illustrates that, even though less in magnitude compared to the radial velocity, the expanding condensate indeed has a finite azimuthal component of the velocity which changes with its anisotropy and thus, demonstrating the breaking of the irrotationality condition in such condensates.

In the first part of the current work, starting from the spinorial version of a time-dependent Gross-Pitaevskii equation (TDGPE) for a two-component SOC-BEC, we 
show in the hydrodynamic limit, with $\bar{s}_{z} \text{(background polarization density)}$ $\ll \bar{n}_{d} \text{(background total density)}$, the corresponding phonon field is described by
\begin{equation}
\frac{1}{\sqrt{-g}}\partial_{\mu}( \sqrt{-g} g^{\mu\nu}\partial_{\nu} \tilde{\theta}_{d})=0 \label{phononeq}
 \end{equation}
Eq.(\ref{phononeq}) is again a  massless scalar field equation in an analogue curved space-time, similar to the Eq. (\ref{phononeqs}) for a scalar BEC, 
but with a fundamentally different $g^{\mu \nu}$.

In the subsequent part of this work, we numerically integrate the TDGPE over a substantial time to directly demonstrate the formation of the sonic
event horizons, the corresponding modulation of the superfluid density as the sonic black hole is formed and then discuss the spontaneous analogue Hawking radiation emitted initially from such SBH. The space-time analogy discussed in the preceding paragraph is valid only in the hydrodynamic regime when the stationarity of the flow is present, and the perturbations in density relative to the background density are small. By studying the time evolution for long-enough time, we show that the dynamics of the configuration inside the SBH at the later times in the simulation do not remain in such a regime due to the black hole lasing phenomena \cite{Steinhauer14,bhlaser}. 
Our study of long-enough time evolution enables us to identify the regime where the sonic analogy for space-time is valid,  and where it breaks down.

However, the regime where the fluctuations inside the SBH are large enough to break the sonic analogy with the real black holes is still interesting to study \cite{sd,Unruh2,Parentani2014}. 
We, in particular, show that at later times, when the sonic analogy is not valid inside the SBH due to the black hole lasing mechanism, the emitted radiation from the SBH is not thermal. To study such a regime, we use the truncated Wigner approximation (TWA) that includes the leading order quantum corrections to the mean-field (GPE) evolution \cite{TWA_fo,TWAbook}. The incorporation of quantum fluctuations can be well approximated, through a stochastic sampling of a Wigner distribution \cite{blakie}, for the system's initial state.  The stochastic initial state is then further propagated through the time-dependent GP equation.

Using TWA, we study in detail the density-density correlation function over a substantial time interval to understand the nature of the emitted radiation from such SBH in $2+1$ dimension created out of this time evolving SOC-BEC. Our analysis demonstrates the occurrence of the amplifying density modulation in such evolving 2+1 dimensional SBH, which was already observed experimentally in the case of a $1+1$ dimensional model of sonic black hole \cite{Steinhauer14}. It also shows, how the nature of the analogue Hawking radiation changes from a spontaneous to the stimulated one over a longer time scale. Finally, we conclude through the illustration of the thermal nature of the radiation at the initial times and the spectral deviation from thermality at later times.  

Accordingly, the rest of the paper is organised as follows. In section \ref{Model} and \ref{metrics}, starting from a TDGPE of a two-component SOC-BEC, using the hydrodynamic approximation, we derive the field equations that describe the phonons corresponding to the total density
field of such two-component SOC condensate. We also discuss the properties of the analogue space-time metric and its connection with the space-time metric of rotating $2+1$ dimensional \cite{Kerr, BTZ,BTZ93} black hole. In section \ref{velcond}, we study in detail the time evolution of the condensate density as it accelerates through a two-dimensional version of time-dependent waterfall potential (Fig. \ref{schematic}) by directly integrating the TDGPE using a split-step method over a substantially long time interval. We also perform a windowed Fourier transformation on the time-evolved condensate density to understand the local features, identify the sonic horizon formation, discuss the nature of radial and azimuthal flow of the evolving density and their consequences. We here have also mentioned a few aspects of the black hole laser phenomena shown by this system. In section \ref{TWA}, we study the density-density correlation function at different times using TWA to understand the nature of the spontaneous analogue Hawking radiation emitted from such a sonic black hole. And, identify the transition from spontaneous to stimulated analogue Hawking radiation with the formation of two horizons. In the subsequent section \ref{SOR}, we study the nature of the radiation spectrum using the density-density correlation function, and then we conclude.

\section{Model}\label{Model} 
To characterise such SBH, we need to calculate the phonon (sound) velocity in SOC-BEC and the condensate velocity that violates the irrotationality condition. Therefore, in this section, we describe the system under consideration and briefly outline how these physical quantities can be evaluated (with their details in the Appendix). We start with the following spinorial time-dependent Gross-Pitaevskii (GP) equation:
\bea 
 i\hbar\frac{\partial \psi_{\kappa}  }{\partial t} & = & \bigg[\frac{\hbar^2}{2m}(-i\partial_x- \kappa\frac{m \eta}{\hbar})^2-\frac{\hbar^2}{2m_y}\partial_y^2 + V_{2D}(\bs{r},t) \bigg]\psi_{\kappa}  \nn \\
 & \mbox{+} &  g_{2D}\bigg( |\psi_+|^2+|\psi_-|^2\bigg)\psi_{\kappa},~ \kappa=\pm  \label{gpe} 
\eea
Eq.(\ref{gpe}) is derived by considering $^{87}$Rb atoms with spin-orbit interactions, where the single-particle Hamiltonian possesses a non-abelian gauge potential of the form $\bs{A}$ : $\it{m}(\eta \check{\sigma}_y,\eta'\check{\sigma}_z,0$) (details in Appendix \ref{1a}), and is given as: 
\bea
\hat{h}&=&\frac{\bs{p}^2}{2 m}\check{I}-\eta {p}_x \check{\sigma}_y-\eta' {p}_y \check{\sigma}_z
\eea
 with $\bs{p}=\{p_{x},p_{y},p_{z}\}$; $\check{\sigma}$ and $ \check{I}$'s are Pauli and Identity matrices respectively and $\eta$,$\eta'$ are the strengths of SO coupling with $\eta'<\eta$. In writing Eq.(\ref{gpe}), we have considered the trapping frequency $\omega_z >> \omega_x,\omega_y$ $(\omega_x=\omega_y)$ and therefore, the dynamics along the z-direction are frozen. 
Here, ${\bs r}$ = (x,y), $m_y = \frac{m}{(1-(\frac{\eta'}{\eta})^2)}$ is the effective mass along y
direction,  $g_{2D}= \frac{V_{int}}{\sqrt{2\pi}a_{\perp}} , V_{int}=\frac{4\pi \hbar^2 a_s}{m}$, $a_s$ is the interatomic scattering length, $a_\perp = \sqrt{\frac{\hbar}{m\omega_z}}$ is the transverse harmonic oscillator length. Also, we have considered the inter and intra-species interaction strengths to be equal \cite{Stringari12}. The external potential $V_{2D}$ [shown in Fig.\ref{schematic}(a)] includes the harmonic confinement. 

Despite the fact that the trap is isotropic, the condensate is not isotropic because of the different effective masses, and the effective momentum along x and y directions [see Eq. (\ref{gpe})]. The appearance of the anisotropic masses is due to the fact that we have considered unequal SO coupling strengths with the condition, $\eta'<\eta$ and, as a result, the effective dispersion gets modified (please refer Appendix \ref{1b}). The shift in the effective momentum in Eq.(\ref{gpe}) is due to the presence of gauge fields. Denoting the spinor order parameter as $\bs{\Psi}(\bs{r})= [\psi_+ (\bs{r}),\psi_-(\bs{r})]^T$, the components of  current $\bs{j}$ obtained from the GP Eq.(\ref{gpe}) (details in Appendix \ref{2b}) are given by, 
\bea
{j}^x &=& \frac{{\hbar}}{2 i m }\bigg( \bs{\Psi^\dagger} \partial_x \bs{\Psi} - \bs{\Psi^T} \partial_x \bs{\Psi^*} \bigg)-  \eta\bs{\Psi^\dagger}  \check{\sigma}_z \bs{\Psi}\nn\\
{j}^y &=& \frac{{\hbar}}{2 i m_y}\bigg(\bs{\Psi^\dagger} \partial_y \bs{\Psi} - \bs{\Psi^T} \partial_y \bs{\Psi^*} \bigg)\label{vel}
\eea
The expression shows the effect of anisotropy and gauge fields clearly and lets us evaluate the velocity of the condensate to describe the horizon location.  The first term in current components is the conventional superfluid current with different effective masses and the second term in ${j}^x$ corresponds to the gauge part, which occurs as a consequence of SOC present in the system.

In the long-wavelength limit, the sound velocities for density ($_{s}$) and polarisation ($_{z}$) modes can be calculated, as
 $c_{s,z}^{x,y}=\hbar\frac{ d\Omega_{d,z}}{d{p_{x,y}} }\bigg|_{p_{x}, p_{y}  \rightarrow 0}$ from the Bogoliubov dispersion for the total density, $n_d=n_++n_-$ and polarisation density, $s_z=n_+-n_-$ given as, $\hbar \Omega_d = \sqrt{E_{-}(\bs{p})[E_{-}(\bs{p}) +2 g \bar{n}_d]}$ and 
$\hbar \Omega_z = E_{-}(\bs{p})$ respectively, where $E_-({\bs{p}})= \frac{p_{x}^{2}}{2m} +  \frac{p_{y}^{2}}{2m_{y}} +\frac{p_{z}^{2}}{2m}$, $\bar{n}_d$ is the background density (Appendix \ref{2}, \cite{brandon}).
For our case $c_{z}^{x,y}=0$, as inter and intra-species interaction strengths are equal. Therefore, we will only consider the density modes and henceforth, drop the suffix ($_s$) and set $c_{s}^{x,y}=c^{x,y}$.
\section{Hydrodynamic formalism and Analogue SBH metric for $2+1$ dimension}\label{metrics} 
We shall now first derive the Eq. (\ref{phononeq}) from the Eq.(\ref{gpe}) using the  hydrodynamic description of the SOC-BEC.  
To this purpose, using the Madelung representation, we write the wavefunction of the condensate $\psi_\kappa({\bf r},t)=\sqrt{n_{\kappa}({\bf r},t)} e^{i\theta_\kappa({\bf r},t)}$ in terms of density $n_{\kappa}$ and phase $\theta_\kappa$ and ignore the quantum pressure term that contains higher-order derivative in density (Appendix \ref{2b}). For linearising the equations, we consider small fluctuations around the background density's ($\bar{{n}}_\kappa$) and the corresponding phase's $(\bar{{\theta}}_\kappa)$ of the condensed part of the two components:
\bea 
n_\kappa({\bf r},t)&\rightarrow& \bar{{n}}_{\kappa}({\bf r},t)+ \tilde{n}_{\kappa}({\bf r},t), \nn\\
\theta_\kappa({\bf r},t)&\rightarrow& \bar{{\theta}}_{\kappa}({\bf r},t)+ \tilde{\theta}_{\kappa }({\bf r},t)\nn\eea
and, retain the terms only in first-order of fluctuations (see Appendix \ref{2b}). In the limit $ \bar{s}_{z} << \bar{n}_{d}$ \cite{Vorinsoc,SOexpansion,Stringari12, stringarirotation} through a straight-forward but lengthy algebra, the linearised hydrodynamic equations yield Eq.(\ref{phononeq}), a second-order differential equation for 
 $\tilde{\theta}_{d}=\tilde{\theta}_{+}+\tilde{\theta}_{-}$, that describes the phonon field with 
   \beq  g^{\mu\nu} = \frac{m m_y}{{\bar{n}_{d}}^2}
   \begin{bmatrix}
   -1&  -v^x &  -v^y \\
  -v^x & {c^x}^2-{v^x}^2 &-v^x v^y \\
   -v^y & -v^y v^x &{c^y}^2-{v^y}^2\\
   \end{bmatrix} \label{gmat} \eeq 
Here, $\bar{n}_{d }=\bar{n}_{+ }+\bar{n}_{- }$, $\bar{s}_{z }=\bar{n}_{+ }-\bar{n}_{- }$, $c^{y}=\sqrt{\frac{g \bar{{n}}_d}{m_y}} < c{^x}=\sqrt{\frac{g \bar{{n}}_d}{m}}$ and the velocities are given, as
\bea {v^x} &=& \frac{\hbar}{m \bar{n}_{d}}[\bar{n}_{+}\partial_x \bar{\theta}_{+} +\bar{n}_{-}\partial_x \bar{\theta}_{-}] - \frac{\eta \bar{s}_{z}}{\bar{n}_{d}} \nn\\ 
{v^y} &=& \frac{\hbar}{m_y \bar{n}_{d}}[ \bar{n}_{+}\partial_y \bar{\theta}_{+} + \bar{n}_{-}\partial_y \bar{\theta}_{-}]\label{velocities}
\eea
The usual velocity-phase relationship now gets modified due to the presence of the term, `$\frac{\eta \bar{s}_{z}}{\bar{n}_{d}}$' leading to  violation of the irrotationality condition and thus, can impart  angular momentum to the BEC \cite{Vorinsoc,SOexpansion} in the absence of any external rotation. Anisotropy in sound and flow velocities make the metric $g^{\mu \nu}$  different from a scalar condensate \cite{visser11} and 
gives an elliptical event horizon and ergosurface. 


The acoustic metric determines the invariant acoustic interval, $ds^2=g_{\mu\nu} dx^{\mu} dx^{\nu}$ of SBH (for details see Appendix \ref{2b}) in SOC-BEC:
\begin{eqnarray}\label{metric}
ds^2& = &\frac{{\bar{n}_{d}}^2(1+\alpha)}{m m_y {c_s}^2} \bigg[-
\bigg(\frac{{c_s}^2}{(1+\alpha)} -[ {{v}^x}^2+{\frac{{{v}^y}^2}{ \alpha }}]\bigg) dt^2\nn\\
&-&2 \bigg({v^x} dx +\frac{{v}^y dy}{\alpha}\bigg) dt + (dx^2 + \frac{dy^2}{\alpha} )\bigg]
\end{eqnarray}
where $\alpha= 1-(\frac{\eta'}{\eta})^2 $ and $c_s=\sqrt{{c^{x}}^2+{c^{y}}^2}$.
We use (t,R,$\phi$) coordinates to write Eq.(\ref{metric}) in polar variables with {\bf R}=(x,$\frac{y}{\sqrt{\alpha}})$ and,  ${{v}^x}=|v|$ cos$\phi$, ${\frac{{v}^y}{\sqrt{\alpha}}}=|v|$ sin$\phi$. Eq.(\ref{metric}) thus, takes the form:
\begin{eqnarray}\label{metricb}
ds^2& = &\frac{{\bar{n}_{d}}^2(1+\alpha)}{m m_y {c_s}^2} \bigg[-
\bigg(\frac{{c_s}^2}{(1+\alpha)} -[ {v^R}^{\,2}+{v^\phi}^{\,2}]\bigg) dt^2\nn\\
&-&2 \bigg({v^R} dR +R {v}^\phi d\phi\bigg) dt + (dR^2 + R^2 d\phi^2)\bigg]
\end{eqnarray}
For a non-zero $v^\phi$, hence $`dt d\phi$' term, the above line element corresponds to that of a rotating black hole \cite{relBEC}. For Schwarzschild metric, such cross-terms of space and time do not appear. The angular momentum per unit mass observed from the outside of a rotating SBH at a radial distance R, in analogy with the real black hole, is given by $g_{t\phi}$ component and is, $J= v^\phi R$.

The above form of the metric though confirms the existence of rotating sonic black hole (RSBH), is not in the more conventional form \cite{BTZ, BTZrev} of a rotating BH in $2+1$ dimension that is $ds^2 = g_{tt}dt^2+2g_{t\phi}dt d\phi+g_{RR}dR^2+g_{\phi\phi}d\phi^2$. 
The line element of the conventional form of the metric straight-forwardly gives the condition for acoustic event horizon and the acoustic ergosurface. The transformation needed to obtain such a form, however, requires that the flow should be axis-symmetric and stationary in the analogue system. Accordingly, to recast the acoustic line element for a SOC-BEC in a more conventional form \cite{BTZ, BTZrev}, we make the following transformation:  
 \bea 
  dt  &\rightarrow&  dt+\frac{-v^R}{\frac{{c_s}^2}{(1+\alpha)}-{v^R}^2}dR,\label{ttrans}\\
d\phi  &\rightarrow&  d\phi+\frac{-v^R v^{\phi}}{\frac{{c_s}^2}{(1+\alpha)}-{v^R}^2}\frac{dR}{R}\label{phitrans}\eea
Axis-symmetric condition is not satisfied rigorously in our simulations using SOC-BEC. However, for some parameter regime in which we consider the 
current problem the numerical simulation suggests that the velocity component of the SBH in a SOC-BEC has only a  weak $\phi$ dependence  and thus, the deviation from the axis-symmetry is not significant. 
To show that the deviation from axis-symmetry, can be neglected approximately, we note that the azimuthal velocity component relative to the radial component is very small, i.e. $v^{\phi}/v^{R} \ll 1$. Therefore, the second term in Eq. (\ref{phitrans}), $\frac{ v^{\phi}/v^R}{\frac{{c_s}^2}{(1+\alpha){v^R}^2}-1}\frac{dR}{R} \ll 1$ and has a tiny contribution to Eq.(\ref{metricb}). Thus using Eq.(\ref{ttrans}) and Eq.(\ref{phitrans}), Eq.(\ref{metricb}) gets modified to:
\begin{eqnarray}
ds'^2=  &{\bar{n}_{d}}^2&\frac{(1+\alpha)}{m m_y {c_s}^2} \bigg[-\bigg(\frac{{c_s}^2}{(1+\alpha)}-[ {v^R}^{\,2}+{v^\phi}^{\,2}]\bigg)dt^2 \nn \\
&-&2 v^\phi R  dt d\phi+\frac{dR^{\,2}}{\frac{\frac{{c_s}^2}{(1+\alpha)}}{{v^R}^2}-1}+R^2 d\phi^2\bigg]\label{RBH}
\end{eqnarray}

The above  form of the line element in Eq.(\ref{RBH}) is now similar to the more conventional line element of the RBH in 2+1 dimensions \cite{BTZ, BTZrev}. From Eq. (\ref{metricb}) we can now delineate the boundaries of ergo-region and event horizon approximately. The ``Acoustic Ergosurface" and ``Acoustic Horizons" respectively can be defined from 
`$g_{tt}$' and `$g_{RR}$' components in Eq.(\ref{RBH}) which gives, 
\bea
\sqrt{{v^R}^2+{v^{\phi}}^2}&=&\frac{{c_s}}{\sqrt{(1+\alpha)}}\label{ergo1} \\
|v^R|&=&\frac{{c_s}}{\sqrt{(1+\alpha)}}\label{ergo2}
\eea
The condition $\frac{{c_s^2}}{(1+\alpha)}  \leq ({v^R}^{\,2}+{v^\phi}^{\,2})$, defines the ergoregion as $g_{tt}\geq0 $ for such a region.
Also, the non-zero $g_{t\phi}$ component implies a local angular velocity for a non-rotating test particle \cite{LTE, LTE1}, 
$\omega = -\frac{g_{t\phi}}{ g_{\phi\phi}} = - \frac{v^{\phi}}{R}\nn$. 
As mentioned, the present case studied in the system is not strictly axis-symmetric, and our discussion is valid only approximately in the limit ($v^\phi\ll v^R$). However, it is possible to create more coherent and axis-symmetric flows in SOC-BEC by tuning the parameters and type of SOC (for example, refer \cite{stringarirotation}, where equal Rashba-Dresselhaus spin-orbit
coupling, was considered). It requires a very detailed analysis and comparison that is beyond the scope of the current work and will be addressed, in future. 

In the subsequent section, we will solve the GPE, Eq.(\ref{gpe}) to see the evolution of the SBH. Ensuring a stationary configuration in SBH, for a time-dependent potential, means there exist a reference frame apart from the lab frame, i.e. a Galilean frame of reference where the potential, $V_{2D}$ and therefore, the mean density of condensed atoms depend only on the space coordinate. To examine whether the requirement of stationary configuration in order to perform the transformation in Eq.(\ref{ttrans}) is being ensured in our simulations, in the next section, we have separated the mean density of the condensed atoms ($\bar{n}_d$) from the time evolved density profile and observed its growth in the Galilean frame which is shown, in Fig. \ref{growth rate}. The space-time analogy and thus, the theoretical mapping to extract the condition for the location of AEH and ergo-region of the SBH, presented in this section, to the system under consideration for the long-time evolution are valid approximately in certain regimes which we will discuss in the next section.
%

\section{time evolution of density}\label{velcond}
In this section, we will study the dynamics of the condensate to get demarcated supersonic and subsonic regions of the sonic black hole. To simulate the SBH, we will use Eq. (\ref{gpe}) and time-evolve the SOC-BEC in a time-dependent 2D potential [see Fig. \ref{schematic}(a)], 
\bea
 V_{2D}(\bs{r},t)=V(\bs{r})+V_{step}(\bs{r},t)\label{pot}
\eea 
where, $V(\bs{r})=\frac{1}{2} m (\omega_x^2 x^2 +\omega_y^2 y^2)$ and $V_{step}(\bs{r},t)= V_s \Theta(r_s(t)-r)$, with $V_s$ ($\sim$ 5$k_B$ nK) is the strength/height of the circular step, $k_B$ is the Boltzmann constant. $r_s(t)=-v_{s}t+r_{s}(t_0)$ is its instantaneous position, where $v_{s}$ ($\sim$ 0.21 mm $s^{-1}$) is the constant speed with which width w(t) decreases, and $r_{s}(t_0)$ is the initial position. $V_{2D}(\bs{r},t)$ accelerates the 2D condensate and, can be compared, to the waterfall potential that simulates $1+1$ dimensional SBH in a recent experiment with scalar condensate \cite{la, Steinhauer14, Steinhauer16}. This potential can be experimentally realised, with the currently available masking techniques \cite{mask}, where, the dynamical potential to be experienced by the atoms, is written on a digital micromirror device (DMD)\cite{Eckel}. The harmonic trap parameters considered for our simulations
are $\omega_x=\omega_y=2\pi \times $ 4.5 Hz, $\omega_z=2\pi \times $ 123 Hz with number of atoms, N$\sim$6000,
and characteristic length, $x_s$=3.41$\mu$m.

\begin{figure}
\subfloat{\includegraphics[scale =0.88]{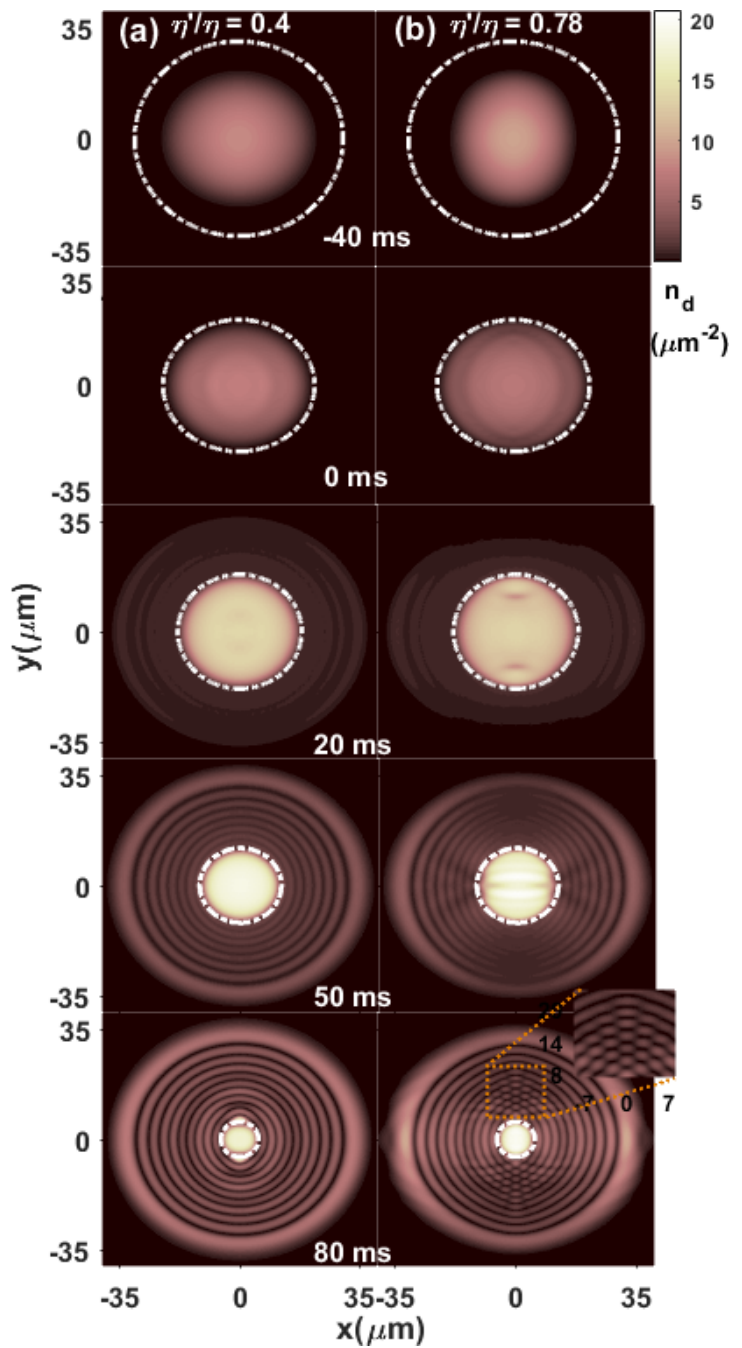}}
\caption{{\it (color online).}Total density evolution for various times corresponding to $\eta'/\eta$ = 0.4 (left), 0.78 (right) is shown. The white dashed lines in each figure represents the location of the circular step.}  \label{MF_GPE}
\end{figure}

We begin with the ground state solution of the GPE Eq.(\ref{gpe}), solved only in the presence of the trap potential, as an initial state for the simulations. It corresponds to the case when the SOC-BEC, is located at the center of the trap, and there is no step potential. In order to accelerate the condensate to supersonic speeds, it is then approached by the step-potential from the outside by decreasing the circular aperture width adiabatically with time. We start the evolution at t = $-40$ ms, when the location of the step is far away from the condensate (Fig.\ref{MF_GPE}). The movement of the step potential towards the condensate, with increasing time, does not impact much on the condensate dynamics till t = $-10$ ms, and significant changes begin to appear in the simulation t = 0 ms onwards. The choice of negative time (also used in \cite{BHL_TETTAMANTI}) indicates time relative to t=0 ms, chosen as the time after which significant changes, was observed in the simulation. As the step potential comes in proximity to the condensate (t = 0 ms), variation in its total density
becomes apparent, and its further long-time evolution, discussed in the following part of the section. 

Time evolution of the total density ($n_d$) for two representative parameter sets of SO coupling strengths $\eta'/\eta =0.4$ and $0.78$, as a function of decreasing w(t) is shown, in Fig. \ref{MF_GPE}. In either case of considered SOC strengths, the density modulation forms a set of concentric ring-like fringes with alternating maxima and minima in an annular region, with increasing anisotropy for a higher ratio of $\frac{\eta'}{\eta}$. The amplitude of the density modulations in this region is small initially and increases with time later. The growth of the density modulation in the bounded annular region, which is the supersonic zone of the SBH, occurs due to the phenomena of ``black hole lasing" (BHL) dynamical instability \cite{bhlaser,Anglin2000}. We will discuss this phenomenon and demonstrate the formation of the supersonic zone later in this section. The variation in the total density in the supersonic region exhibits modulations, and thus a localised behaviour. Therefore, we will carry out a local Fourier transform - the windowed Fourier transform (WFT) \cite{WFTbook,jacobson17} on the total density in the neighborhood of a given position, by filtering it with a window, to understand the amplification of time-dependent density profile of the BEC. It is needed to extract local spectral information and hence, for separating the background condensate density ($k$=0) from the oscillatory spatial density (with $k \neq$0).

\begin{figure}[htpb]
\includegraphics[scale =0.65 ]{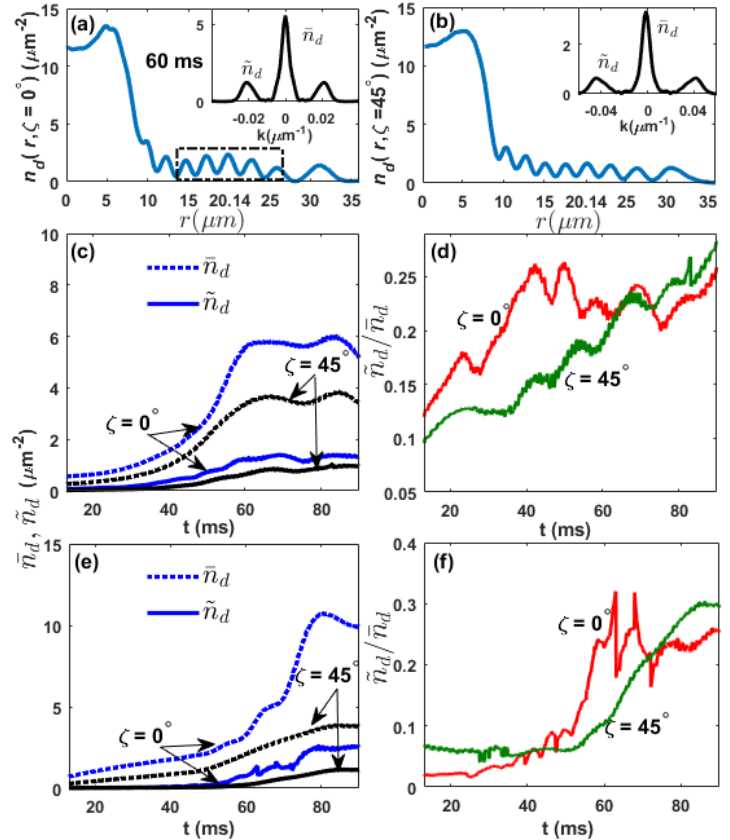}
\caption{{\it (color online).} (a-b) shows the cross-sectional total density at 60 ms, along $\zeta= 0^\circ, 45^\circ$ for $\eta'/\eta = 0.4$ respectively. Inset shows 1D-WFT, $n_d(k,r)$ with a window of width D = 6.5 $\mu$m, centred at r = 20.14 $\mu$m. (c) Growth of background density, $\bar{n}_d$ and the fluctuation, $\tilde{n}_d$ as a function of time and, (d)  shows the ratio of fluctuations in total density to background total density as a function of time. (e,f) shows the results corresponding to (c,d) respectively for the case $\eta'/\eta = 0.78$. }\label{growth rate}
\end{figure}

To this purpose, we will take cross-sectional density along a certain azimuthal direction, $\zeta$ and then utilise the one dimensional-WFT of the total density, which is defined, as:
\bea
n_d(k,r) = \int_{-\infty}^{\infty}  dr' n_d(r',\zeta) w(r'-r;D)e^{-ikr'}\label{denswft}
\eea
where $w(r'-r;D) = \frac{1}{\sqrt{\pi} D} exp[-(r'-r)^2/D^2]$ is a Gaussian window of width D. Eq.(\ref{denswft}) constitutes a local Fourier transform of the GP density, $n_d$ and captures features that vary on length scales $\ll$ D.
 We have shown the cross-sectional density, $n_d$ for two representative angles $\zeta = 0^\circ, 45^\circ$ corresponding to $\eta'/\eta = 0.4$, in Fig. 
\ref{growth rate} (a,b). The corresponding 1D-WFT is shown, in their insets at 60 ms, which have a peak at k=0 ($\bar{n}_d$) that represents the condensed part of density, and two $ k \neq 0$ peaks ($\tilde{n}_d$) representing the non-condensed part of  density. Thus, by evaluating the WFT of density through Eq. \ref{denswft}, we  extracted the density amplitude at $k$ = 0 and $k$ $\neq$ 0 and, observed the growth of these at $r$ = 20.14 $\mu$m as a function of time which is shown, in Fig. \ref{growth rate} (c) and for $\eta'/\eta = 0.78$ is shown, in Fig. \ref{growth rate} (e). Here the size of the window, D = 6.5$\mu$m is chosen, in a way that it covers the local variations on length scales $\ll$ D precisely. The growth of condensed as well as non-condensed density is exponential till $\sim$ 60 ms for $\eta'/\eta = 0.4$, and approximately $\sim$ 75 ms for $\eta'/\eta = 0.78$; after that, the increase in the growth saturates randomly. The ratio between the non-condensed and condensed part of the  density, for both the cases, as a function of time is shown, in Fig. \ref{growth rate}(d,f). 

The space-time analogy discussed in section \ref{metrics} is valid only in the hydrodynamic regime when the fluctuations are small ($\tilde{n}_d$/$\bar{n}_d \ll$ 1), and the condensate flow is stationary. Stationarity in SBH configuration means that the mean density should remain independent of time in the Galilean frame of reference. It is required to validate the transformation performed in Eq.(\ref{ttrans}), used for theoretical mapping of wave equations for the phonons in condensate flow to that of a massless field in curved space-time (section \ref{metrics}). From Fig. \ref{growth rate} (c,e), we observe that the growth of the mean density of condensate is very small up to the initial time
 $\sim$ 30 ms for $\eta'/\eta = 0.4$ and approximately the same up to $\sim$ 50 ms for $\eta'/\eta = 0.78$.  After that,
the growth is considerable and the break down of the space-time analogy, inside SBH, after these times particularly owes to non-stationarity of the flow. At later times, after $\sim$ 60 ms for $\eta'/\eta = 0.4$ and approximately $\sim$ 75 ms for $\eta'/\eta = 0.78$ the growth
approximately saturates but, due to the large density fluctuations, the sonic analogy breaks inside SBH. Thus, the space-time analogy of our SBH model, inside the supersonic region, is valid approximately up to $\sim$ 30 ms for $\eta'/\eta = 0.4$, 
and $\sim$ 50 ms for $\eta'/\eta = 0.78$. However, in the outside region A of SBH [Fig. \ref{schematic} (b)], the background density is approximately stationary (Fig. \ref{MF_GPE}) and also, the density fluctuations relative to the background density is negligible. Thus in our simulation, the space-time analogy is preserved in the outside region A nearly at all the times, but up to certain times in the inside region B of SBH.

The density modulation is mostly radial (along the radius) for the values chosen in our simulation. But for the higher ratio of anisotropic strengths, the later time simulation [$60-80$ ms, Fig.\ref{MF_GPE}(b)] shows the formation of the interference fringes along the azimuthal direction $\zeta$, for a fixed radius, as well [inset, Fig.\ref{MF_GPE}(b) at $80$ ms]. Particularly, density in Fig.\ref{MF_GPE}(b) at $80$ ms, shows density minima where it vanishes along the azimuthal direction (inset). At this time, the flow becomes approximately stationary as the growth of background density nearly saturates [Fig. \ref{growth rate} (e)], but the space-time analogy breaks down due to large fluctuations relative to the background flow. Studying this regime might not answer about the scenario in real black holes, but can provide a few traits of the long-term behaviour of black hole laser flows \cite{Parentani15,Anglin}.

 Thus, to understand these density minima along the azimuthal direction better, we plot its corresponding phase variation in Fig. \ref{phase} (Appendix \ref{2b}) and identify that these minima's correspond to abrupt changes in the phase like the one that appears when vortices/phase slips are present in the condensate \cite{stringarirotation,buttsrokshar}. We, therefore, observe the presence of vortices in the supersonic region, at the later times in the simulation. This observation is also consistent with the long-term behaviour in similar transonic bounded flows, which leads to the formation of solitons in one-spatial dimension \cite{Parentani15} and vortices in spatially two-dimensional model systems \cite{Anglin} and, followed usually by their emission in the neighbouring subsonic region. In our case, due to the inward movement of the step potential at later times in the simulation ($\sim$ 90 ms) and BHL instabilities, the supersonic region becomes unstable, and consequently eventually disappears at following times. Thus, we restrict the study of dynamics in our simulation for times up to $\sim$ 90 ms.
\begin{figure*}
\includegraphics[scale=0.62]{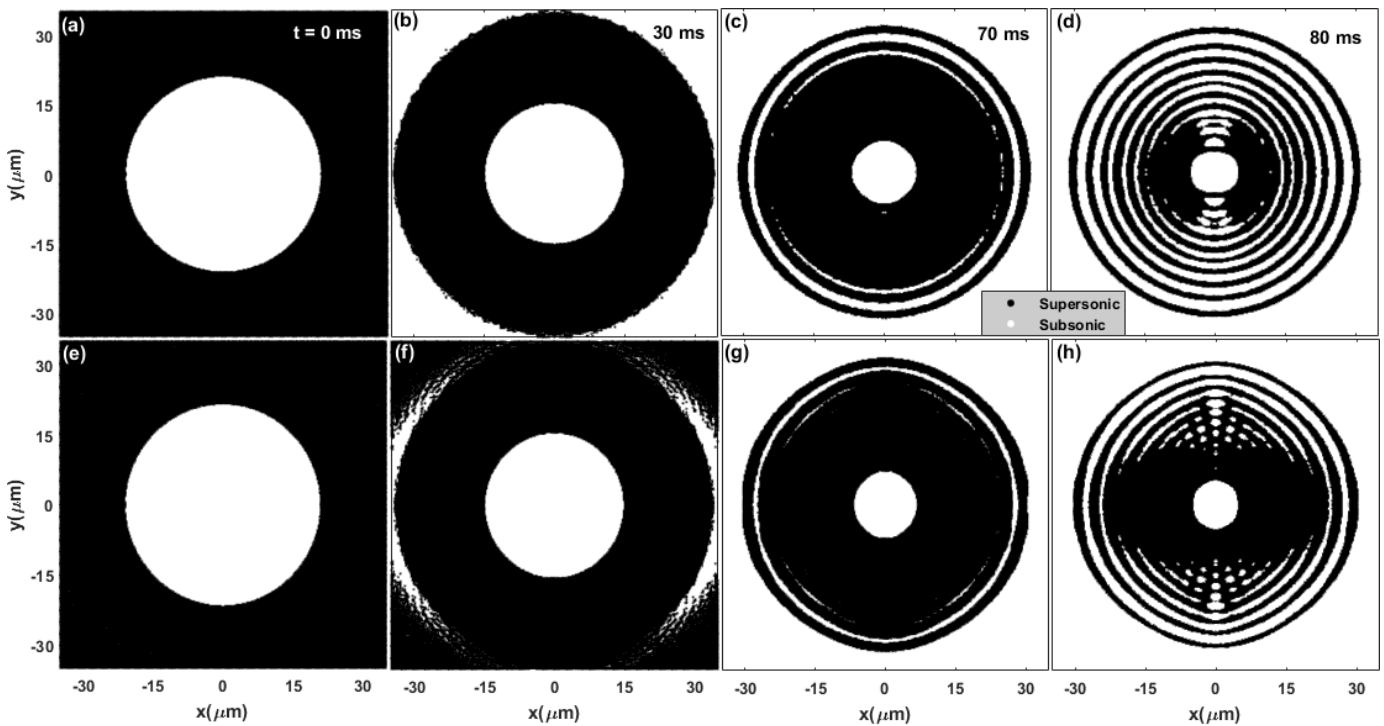}\\
\centering
\caption{{\it (color online).} {\it Evolution of the SBH:} The figure shows the formation of a single horizon (AEH) of the SBH at the initial time and, as time progresses IH also gets formed, which is followed by, the stratified formation of ``local" supersonic-subsonic layers in the annular region. The evolution is shown in (a-d) for $\eta'/\eta$ = 0.4 and in (e-h) for $\eta'/\eta$ = 0.78, at various times: 0,30,70 and 80 ms respectively. Data beyond IH layer, after its formation, is not shown.} \label{vflow2}
\end{figure*}

 To investigate further regions of subsonic, i.e. $|v^R| < \frac{{c_s}}{\sqrt{(1+\alpha)}} $ and supersonic velocity, i.e. $|v^R| > \frac{{c_s}}{\sqrt{(1+\alpha)}} $ are shown, in Fig.\ref{vflow2}. At t=0 ms, when step potential location is in the vicinity of the condensate, we get distinguished supersonic (black) and subsonic (white) region and thus, the SBH gets formed. At this time, one of the horizons, acoustic event horizon (AEH) [using Eq.(\ref{ergo2})] of the SBH gets formed approximately at the step-potential location. This is shown in Fig.\ref{vflow2}(a,e) for two parameter sets $\eta'/\eta =0.4, 0.78$ respectively. As the time evolves, another closed boundary - inner horizon (IH) gets formed. It is formed at 30 ms for $\eta'/\eta = 0.4$ and is shown, in Fig. \ref{vflow2}(b). For $\eta'/\eta = 0.78$, the closed IH is formed later at 50 ms (we have not shown here). We observe a stratified structure of subsonic and supersonic zones at later times in our simulation, in the annular region, apart from the outermost and innermost boundaries, shown in Fig.\ref{vflow2}(c-d,g-h). It occurs due to the large amplitude modulation inside the SBH, between the two boundaries - AEH, IH and is another related consequence of BHL instability at the later times in our simulation, where the space-time analogy no longer holds. Also, we have shown the result in Fig. \ref{vflow2} (b-d,g-h) only till IH formation, as our analysis is confined, up to this region. The AEH and IH of the SBH at later times in this work refer to the boundaries of the earlier black hole, where the sonic analogy was valid. This terminology at later times, is used, to delineate the boundaries of supersonic-subsonic regions in the SBH.

A conventional $1+1$ dimensional SBH \cite{TWA2008} usually have a delineated division of supersonic and subsonic regions. Recent experiments on $1+1$ dimensional SBH formation using BEC \cite{Steinhauer14,Steinhauer16}, in a similar set-up, however, observes few local intersections in the supersonic region at later times which corresponded to points  \cite{jacobson17,Steintheory,BHL_TETTAMANTI, BHL_Plata}. It is because of the presence of ``bounded" supersonic zone in such configurations. Due to the two-dimensional nature of this work, we can here clearly visualise the formation of closed separated regions due to such intersections. These are present due to the large amplification of modulated density pattern in the supersonic region at later times, which creates small local supersonic and subsonic division (following the same definition used for defining these regions).   
 
In our case, as the condensate is located at the centre of the trap and is approached by the step potential from outside, AEH and IH lie respectively at the inner and outer-boundary [shown in Fig.\ref{schematic}(b)], and the flow is towards IH. The AEH is formed approximately at the step potential location; IH is created, at the boundary, where particles get reflected.
 The convention used to define AEH and IH, here for the two-dimensional SBH, is also in consonance with the typical experimental configuration of the analogue one-dimensional black hole\cite{Steinhauer14,Steinhauer20} and related numerical works \cite{jacobson17,jacobsoncorr,Steintheory}. The emission of radiation from the AEH for such an SBH is illustrated, in Fig. \ref{schematic}(b) and Fig. \ref{Spacetime} (top) (in Appendix \ref{2b}). The emission takes place towards the central disk, i.e. in the inward direction (as viewed from outside). However, the direction of the emitted radiation in a 1D configuration with two horizons is outwards \cite{Steinhauer14,bhlaser}. It is due to the type of potential profile chosen for this two-dimensional SBH formation, where the step is approaching the condensate from outside.

 \begin{figure*}
\includegraphics[scale=0.68]{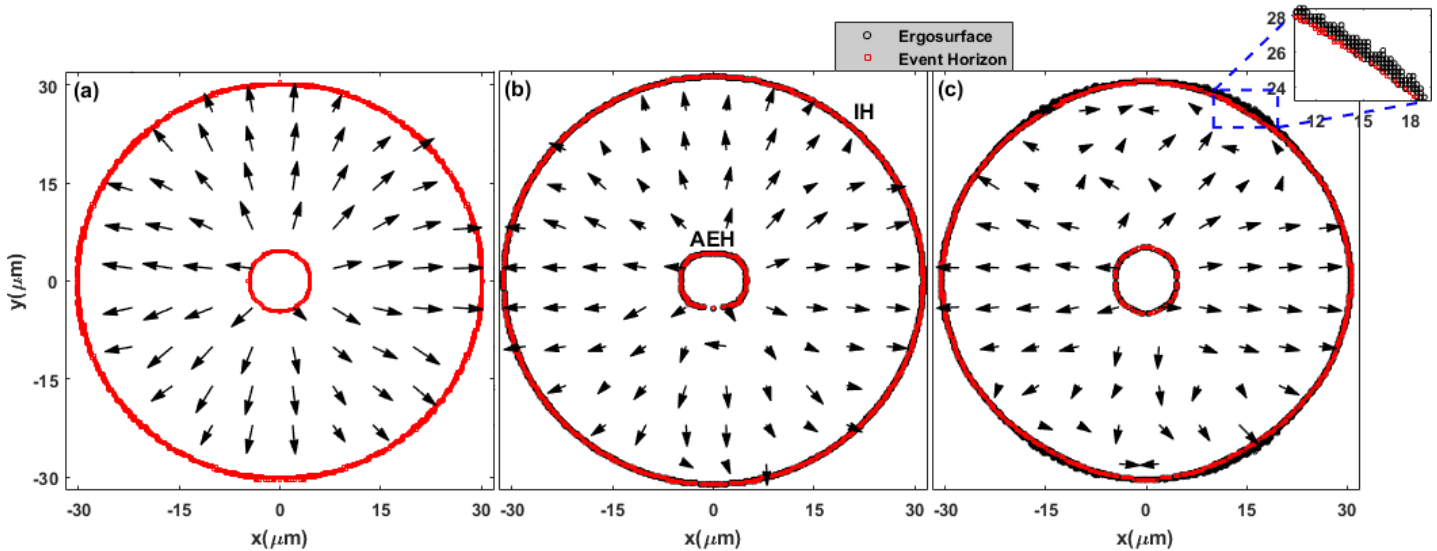}\\
\centering
\caption{{\it (color online).}(a-c) {\it Comparison of flow in a 2D two-component BEC (without spin-orbit coupling) to the SOC case for the same set-up considered in this work} : (a) Normalised velocity ($v_x/|v|,v_y/|v|$) at 80 ms, in the supersonic region, for a two component BEC without spin-orbit coupling. (b,c) To illustrate the azimuthal flow in a SOC-BEC, we plot the normalised velocity for $\eta'/\eta$ = 0.4, 0.78 respectively at 80 ms. Length of arrows represents the magnitude of the velocity. The figures are superposed with the boundaries of the event horizon(red) and ergorsurface(black) of the SBH. Inset in (c), shows slight distinction between the two. }\label{vflow1}
\end{figure*}
Near the end of our simulations, the growth of fluctuations in the density nearly saturates, and the flow becomes quasi-stationary inside the SBH (Fig. \ref{growth rate}). Although the space-time analogy is not valid at later times, to visualise the flow at such times and to illustrate the irrotationality aspect in SOC-BEC, we plot the components of normalised velocity vectors ($v_x/|v|,v_y/|v|$) in the supersonic region, superposed with the ergosurface and event horizon in Fig. \ref{vflow1}. As a comparison, a two-component BEC, i.e., the case without spin-orbit coupling is time evolved in the same potential, Eq.(\ref{pot}) with equal interaction strengths (= $g_{2D}$). The generated flow, shown in Fig. \ref{vflow1}(a), is almost radial. The density evolution corresponding to this case, shown in Fig. \ref{scalar_te} [details at the end of Appendix \ref{2b}].

As we include the spin-orbit coupling in the system, we see a finite but small azimuthal component present locally in the flow. It can be seen, at some places in Fig. \ref{vflow1}(b) for $\eta'/\eta$=0.4 (small ratio). For a slightly higher ratio ($\eta'/\eta$=0.78), the rotational character in the local flow, characterised by a finite azimuthal component of velocity vectors in the flow, increases. The irrotationality condition, as compared to scalar BEC, thus breaks down in SBH formed using a SOC-BEC. For a two-component BEC without SOC placed under similar condition, the event horizon and ergo-region boundary exactly coincide. For the present study of SBH from SOC-BEC, $v^{\phi}/v^{R} \ll 1$, thus these two boundaries  almost coincide for $\eta'/\eta = 0.4$, within our numerical accuracy. However, a closer inspection of Fig.\ref{vflow1}(c, inset), for $\eta'/\eta = 0.78$, shows the presence of azimuthal flow (i.e. along $\hat{\phi}$) which leads to a slight separation between the two boundaries, indicating a relatively larger azimuthal velocity.

The amount of rotation in the supersonic region of such SBH formed out of SOC-BEC is small, and this can be explained, through Eq. (\ref{velocities}). To increase the rotation coherently, one needs to increase $\phi$ = $\tan^{-1}(v^y/v^x)$ smoothly by changing the ratio $v^y/v^x$. It can be achieved, by increasing $v_{y}$, which is nevertheless constrained by the limit $\eta'/\eta < 1$. The effect of varying $\eta'€™/\eta$ is visible through the comparison of Fig. \ref{vflow1} (b) and (c). 
But for a fixed SOC strength, $\phi$ can be also be increased by decreasing $v_{x}$.
Now, here we have considered relatively less polarization density (background) as compared to the total density (background), i.e. $\bar{s}_{z} << \bar{n}_{d}$ under which the Eq. (\ref{phononeq}) etc. are derived. Thus, the second term in $v^x$ [in Eq. (\ref{velocities})] that depends on the ratio $\bar{s}_{z}/\bar{n}_{d}$ has a tiny (but non-zero) contribution to the angular variation $\phi$ in comparison to the first term.

It may be noted that the circulation of the velocity field, $\textbf{\emph{v}}$ over a closed contour C given by the line integral $\oint_C \textbf{\emph{v}} \cdot \,d{\bf l}$, gives the measure of rotation. Here, $d{\bf l}$ is the line element on the curve C.
For a coherent flow, along the contour C, the major part of of the velocity field should be tangent to the curve for contributing maximally to the line integral mentioned earlier (with the same sign). The coherent flow in the analysis discussed here is, however, turning out to be small. The potential in Eq. (\ref{pot}), is generating a strong radial flow in the supersonic zone due to the movement of the circular step potential, $V_{step}(\bs{r},t)$. Thus the addition of this potential also cannot induce any coherent rotation to the time evolved condensate.
  \begin{figure*}
\subfloat{\includegraphics[scale=1.1]{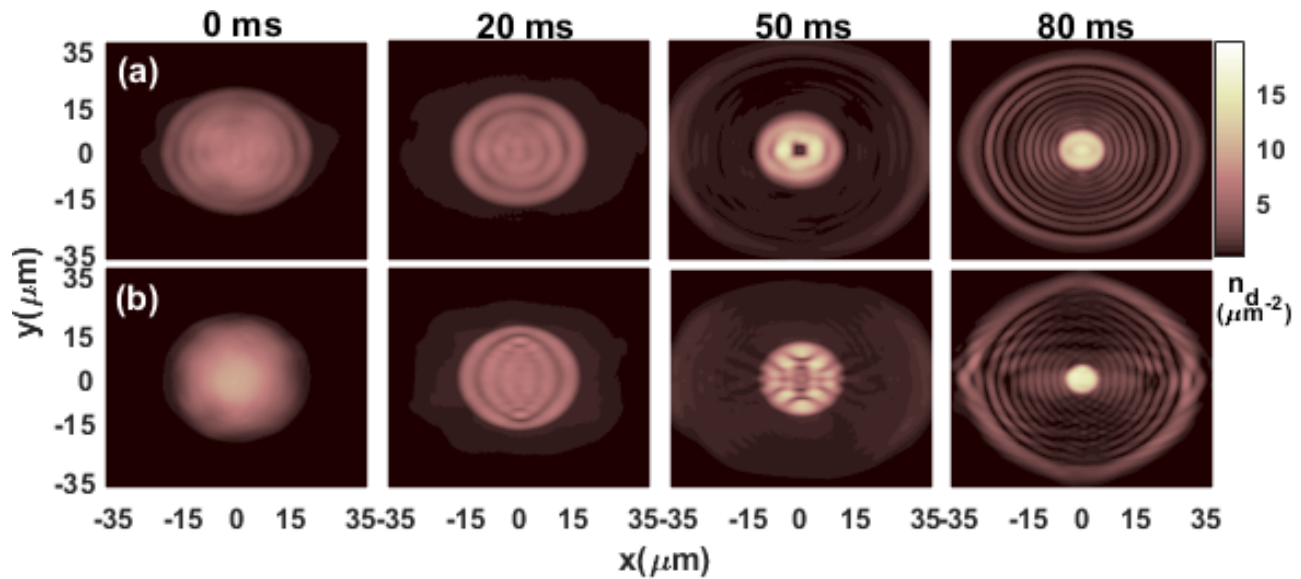}}\\
\caption{{\it (color online).} Time evolution of total density for one of the realisations from the ensemble using TWA,
at various times for $\eta'/\eta =$ (a) 0.4 and (b) 0.78.}\label{TWA_GPE}
\end{figure*}

In this set-up for SBH simulation, AEH formation is followed, by the creation of IH, and the amplification of the density modulations occur in the region bounded by AEH and IH. The presence of these two horizons, the exponential growth of the density pattern approximately up to $\sim$ 60 ms for $\eta'/\eta = 0.4$, Fig. \ref{growth rate}(c) and the superluminal dispersion (shown in Fig. \ref{sup_disp}) suggests that the system exhibits black-hole lasing phenomena following the theory proposed by Corley and Jacobson \cite{bhlaser}. The SBH configuration considered in this work consists of a single horizon (AEH) at the initial times and two closed horizons (AEH and IH) - from $\sim$ 30 ms onwards for $\eta'/\eta = 0.4$ and $\sim$ 50 ms onwards for $\eta'/\eta = 0.78$. Inhomogenous flows in sonic black holes with a single horizon can be dynamically stable but, are always energetically unstable \cite{Parentani15,Parentani2013}. Dynamically stability, that can be, identified using the Bogoliubov–de-Gennes approach implies the non-existence of any complex eigenfrequency, which can ultimately result in the growth of the modes. Such configurations are energetically unstable because linear perturbations with negative energy always exist in the supersonic region \cite{Parentani15}. 

Additionally, mixing of these modes with the external, outgoing positive energy modes, i.e. in region A leads to spontaneous pair-production of phonons with opposite energies through which analogue Hawking radiation is observed, in such SBH \cite{bhlaser}. The phenomenon of this mechanism is yet more abundant when the SBH configuration has a finite supersonic region, bounded by AEH and IH. The negative energy partners of Hawking radiation striking the second horizon (inner horizon) will get reflected towards AEH, and these reflected modes stimulate further Hawking radiation repeatedly resulting in black-hole laser mechanism \cite{bhlaser}. This phenomenon constitutes a dynamical instability that involves quasi-particle pair-creation \cite{bhlaser,Anglin2000} and then leads to the exponential amplification of the non-condensed part of the density up to a specific time, as mentioned earlier [Fig. \ref{growth rate} (c,e)]. Thus, due to the mechanism of BHL dynamical instability in this SBH configuration, the sonic analogy in our simulation at later times breaks down.

The amplification of density modulations in the annular region is reminiscent of the similar amplification process in quasi-one-dimensional SBH in recent experimental work on the formation of analogue black hole laser and associated theoretical works \cite{la,Steinhauer14,Steinhauer16,jacobson17,jacobsoncorr,Steintheory}. However, there is a divergent view on the role of Hawking radiation to this amplification process. There are works in literature that support \cite{Steinhauer14,Steintheory,BHL_TETTAMANTI,Parentani15} and contradict \cite{jacobson17,jacobsoncorr} its role. In ref.\cite{Steinhauer14,Steintheory} the growth of density modulations, between inner and acoustic event horizon, were interpreted as self-amplifying Hawking radiation. However, some theoretical works that studied these experiments \cite{jacobson17,jacobsoncorr} relate the amplification of density modulation to Bogoliubov-Cerenkov radiation (BCR) mechanism. In a recent preprint \cite{Steinhauer20}, this amplification process shows that both of these phenomena are present in the amplification of density modulations in the supersonic region with one dominating over the other. We will not go into this discussion and postpone its clear role, for this system, to the amplification process for some future work.

In the subsequent section, we will evaluate the density-density correlation function corresponding to the time evolved density profile of the condensate by using the truncated Wigner approximation (TWA) \cite{TWA2002,TWAbook,TWA2008}. The approximation involves the addition of initial distribution of random fluctuations in the Bogoliubov modes, whose coefficients are randomly populated according to Gaussian distribution with zero-mean, to the GP wave function to create an ensemble of various realisations which is followed by an ensemble averaging. We will analyse the radiation emitted from this set-up using the density-density correlation function and, show that the initial radiation emitted from this set up is analogue Hawking radiation in the following sections.

\section{Correlation function} \label{TWA}
\begin{figure*}
\includegraphics[scale=0.9]{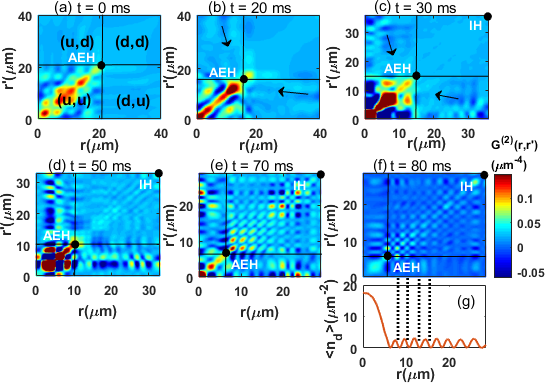}\\
\medskip
\vhrulefill{1pt}\par
\vspace{15pt}
\includegraphics[scale=0.9]{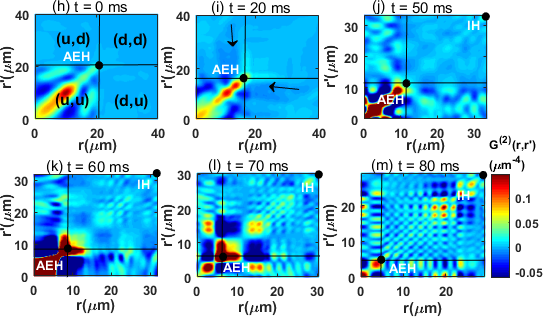}\\
\caption{{\it (color online).} Density-density correlation, $G^{(2)}(r,r';0^\circ)$ obtained using TWA, for $\eta'/\eta$ =  0.4 and 0.78 is shown in (a-f) and (h-m) respectively at various times mentioned on the top of each figure. (g) shows the ensemble-averaged cross-sectional density, $\langle n_d(r,0^\circ) \rangle$. Arrows marked in (b,c; i), along the negative correlation band emanating from the AEH, are indicative of the presence of spontaneous Hawking radiation emitted from AEH as discussed in the text. AEH and IH, are marked for all the illustrative times.}\label{TWAzeta0}
\end{figure*}
To extract the information about the radiation emitted from this SBH, we use the density-density correlation function, which is a powerful experimental tool to detect analogue Hawking radiation from SBH \cite{jacobsoncorr,TWA2008, Steintheory}, 
\bea 
G^{(2)}(r,r';\zeta)= \langle \delta n_d(r,\zeta)  \delta n_d(r',\zeta)\rangle \label{CF}
\eea
where $G^{(2)}(r,r';\zeta)$ is the two-point equal-time connected density-density correlation function
along different azimuthal directions, 
$\delta n_d = {n}_d - \langle {n}_d \rangle$ is the fluctuation in the density. $\zeta$ is the angle in the x-y plane and $r$, $r'$ correspond to the radial coordinate inside and outside the acoustic event horizon. The expectation values ($\langle ... \rangle$) in Eq.(\ref{CF}) are computed by taking an ensemble $\sim$ 100 of BEC. 
For creating an ensemble with several realisations to compute the correlation function, we shall adopt the truncated Wigner approximation (TWA) \cite{TWA2002,TWAbook,TWA2008}. 

To account for quantum fluctuations, in this approximation,  one adds the fluctuation term (noise) of the ``linear order", $\delta\psi_\kappa$ in each component of the GP wave function which models small excitations on a stationary zero-temperature condensate, $\psi_{\kappa 0}({\bf r})$. But, the resulting stochastic wave function of each component is evolved however classically, i.e. by using the mean-field description of BEC, through GPE, Eq.(\ref{gpe}). The resulting stochastic wave function for a single realisation, of an individual component, taken at the beginning of time evolution is:

\bea
\psi_\kappa({\bf r}) = \alpha_0 \psi_{\kappa 0} ({\bf r}) + \sum_j \bigg(\alpha_j u_{\kappa j}({\bf r})  + \alpha_j^* v_{\kappa j}^*({\bf r}) \bigg)
\eea
with $ \delta \psi_\kappa({\bf r}) = \sum_j  \bigg(\alpha_j u_{\kappa j}({\bf r})  + \alpha_j^* v_{\kappa j}^*({\bf r})  \bigg) $ where the index j refers to the sequence of the quasi-particle
excitation (j = 1,2,...,100 in our simulations). The complex functions $u_{ij}$ and $v_{ij}$ denote the Bogoliubov quasi-particle excitations with the normalisation, $\int d{\bf r} \sum_\kappa \big(|u_{\kappa j}|^2 - |v_{\kappa j}|^2 \big) = 1$;
 $\alpha_j$ is a complex random variable with a probability distribution, $
 W(\alpha_j,\alpha^*_j) = \frac{2}{\pi}  \exp[-2|\alpha_j|^2]$ at T=0 K. The total number of atoms ($N= N_c + N'$) is kept fixed, therefore,  $N_c =N-N'$ results in the number of atoms in the condensed state. The coefficient of the condensate component is given by, $\alpha_0 =\sqrt{N_c+\frac{1}{2}}$ and
$ N' = \int d{\bf r} [ \sum_j (|u_{\kappa j}|^2 + |v_{\kappa j}|^2 ) (|\alpha_j|^2 -1/2) +  \sum_j |v_{\kappa j}|^2 ]$ gives the number of excited atoms corresponding to each component. Using TWA, the time evolution corresponding to the total density for one of the realisations from the ensemble for $\eta'/\eta = 0.4$ and $0.78$, is shown in Fig. \ref{TWA_GPE}. A 1D cross-sectional density profile, comparing the simulations of Fig. \ref{MF_GPE}(a) from section \ref{velcond} and Fig. \ref{TWA_GPE}(a), is shown in Fig. \ref{1dtwa_gpe} (Appendix \ref{TSHR}). 


 We will discuss the behaviour of the correlation function at various times for this set-up of SBH to gain insight into the radiation emitted from this configuration, presented in the next section. Figure \ref{TWAzeta0}, shows the density-density correlations between the points in the inside-inside (d,d); outside-outside (u,u) and inside-outside [(d,u);(u,d)] regions of the SBH along $\zeta=0^\circ$, as a function of time for $\eta'/\eta = 0.4,0.78$ in (a-f) and (h-m) respectively. The quadrants are marked, with subsonic upstream region (``u") and supersonic downstream region ("d").
We show here the results from t = 0 ms onwards when the AEH gets formed [Fig. \ref{vflow2}(a,e)], and study the correlations between the points lying outside/inside the SBH in these regions. We will firstly discuss for $\eta'/\eta =0.4$ and then later, for $\eta'/\eta =0.78$.
 At t = 0 ms, the density-density correlation is highest along the diagonal in the (u,u) region because in this part total density is high for the range $0\le r \sim 21$, as can be seen in Fig. \ref{1dtwa_gpe} (Appendix \ref{TSHR}) at the corresponding time.  
We have marked AEH and IH (whichever forms), for all the demonstrated times, in the correlation figure.

After the formation of AEH, as time increases, we observe two bands with negative correlation originating from the AEH (along the arrows marked) in the (u,d) and (d,u) regions at t = 20 ms (also at 10 ms). These bands are related to the Hawking-partner correlations between various points located on opposite sides of the horizon, as each point along this band represents their equal propagation times from the horizon. These correspond to the spontaneous Hawking radiation in literature \cite{Steinhauer16,Steintheory,Steinhauer18}. We have also identified it in the next section. For the spontaneous Hawking radiation, there are no incoming modes from either side of the horizon \cite{Steinhauer18}. As there is no IH formed yet, therefore, there is no possibility of modes coming back. Apart from this, in the diagonal part of (u,u) region, one bright band is accompanied by two dark blue bands, which occurs as a consequence of two dips in the oscillation in the central part of density Fig. \ref{1dtwa_gpe} (Appendix \ref{TSHR}).

At t = 30 ms, the formation of IH is seen, in Fig.  \ref{vflow2}(b). However, at this time, the dark-bands emanating from the AEH are only faintly visible. After this stage, the emission in the (u,d) and (d,u) regions will no longer correspond to spontaneous emission, due to the presence of IH the particles from the IH get reflected towards the AEH. In the (u,d) and (d,u) regions, alternately bright and dark fringes begin to appear.

At t = 50 ms, checkerboard-like pattern, i.e. alternate maxima-minima rectangles (but non-uniform) begin to form in some area of (d,d) region but is very faint. The pattern is smudged near the AEH and IH. 
At t = 70 ms, the checkerboard feature becomes more prominent in (d,d) region, but is not uniform across the whole area. Though it is clear near IH, it remains smeared near AEH at this time. The diagonal along the (d,d) and (u,u) region is different as the density's in both the `u' and `d' regions differ. 
At t = 80 ms, checkerboard feature is very clear near the AEH in (d,d) region. The alternate dark and bright fringes become parallel in the (d,u)/(u,d) region. The periodicity of the modulations in the total density resembles very closely to these fringes, as illustrated in Fig. \ref{TWAzeta0}(g). 

For the higher ratio of the SOC strengths $\eta'/\eta =0.78$, a closed inner horizon is being formed somewhat later at 50 ms as compared to the earlier case with the lower ratio of the SOC strengths, $\eta'/\eta =0.4$. However, in this case with higher anisotropy, the IH begins to form partially among certain directions from $\sim$ 30 ms onwards [Fig. \ref{vflow2}(f)]. Thus, the observation of spontaneous Hawking radiation along all direction persists till $\sim$ 30 ms. The fluctuations in the total density relative to the background density up to $\sim$ 50 ms are small, as shown in Fig.\ref{growth rate}(e). Also, the perturbative modes after reflection from the partially formed IH may escape from the open boundaries in this case. Thus despite a partly formed second boundary (IH) from $\sim$ 30 ms onwards, there is not much change in the (d,d) region till $\sim$ 50 ms. Dark bands emanating from the AEH are still visible but, in this case, we do not call it spontaneous as it could have resulted due to reflection of perturbative modes from any part of partially formed IH boundary. Soon after the closed formation of the IH at $\sim$ 50 ms, a non-uniform checkerboard-like pattern begins to appear near IH in the (d-d) region at 60 ms. This pattern becomes prominent in (d-d) region at 70 ms, but is still non-uniform and remains smudged near AEH.  A clear pattern in (d-d) region, is seen at 80 ms.

To summarise, the following events were observed, in the sequence listed: 
 only one of the horizons is formed (AEH) at initial times (Fig. \ref{vflow2});
spontaneous HR is observed, for times t $\sim$ $10-30$ ms for both the cases;
Closed boundary of IH is formed at t = 30 ms (50 ms for $\eta'/\eta$ = 0.78); The formation of the IH, either partly ($\eta'/\eta$ = 0.78) or completely ($\eta'/\eta$ = 0.4), marks the end of spontaneous stage of emitted analogue Hawking radiation \cite{Steinhauer20}; density modulation begin to form in the supersonic regime. After this stage, the emission in the (u,d) and (d,u) regions is no longer spontaneous and corresponds to stimulated process \cite{Steinhauer16,Steinhauer20} due to the presence of IH. Also, the checkerboard pattern begins to form in the (d,d) region and becomes amplified with time.
\begin{figure}[htpb]
\includegraphics[scale=0.82]{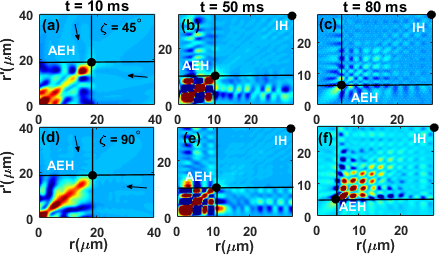}\\
\includegraphics[scale=0.82]{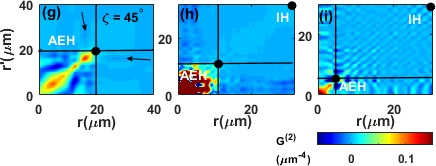}
\caption{{\it (color online).} (a-c; d-f) Density-density correlation for $\eta'/\eta$ =  0.4, along $\zeta = 45^\circ$ and $90^\circ $ respectively, at various times mentioned on the top. (g-i) $G^{(2)}(r,r';45^\circ)$ at these illustrative times for $\eta'/\eta$ =  0.78. Arrows marked in $G^{(2)}$ at the initial times in (a,d; g) are just symbolic of the presence of spontaneous Hawking radiation emitted from the horizon. AEH and IH are also marked.}\label{TWAzeta4590}
\end{figure}

We also observe that the behaviour of the correlation function is direction-dependent. However, the physical phenomena described above remains similar. It is illustrated, in Fig. \ref{TWAzeta4590} along $\zeta = 45^\circ$ and $90^\circ$, for the same set of parameters at some representative times. Also, we have checked that (but not shown here) the behaviour along different cross-sections (along a particular axis) is same, i.e. the behaviour of $G^{(2)}$ at $\zeta = 0^{\circ}$ is identical to $\zeta = 180^{\circ}$; behaviour at $\zeta = 45^{\circ}$ is identical to $\zeta = 135^{\circ}$ and similarly for all the angles.
\begin{figure}[htpb]
\includegraphics[scale=0.9]{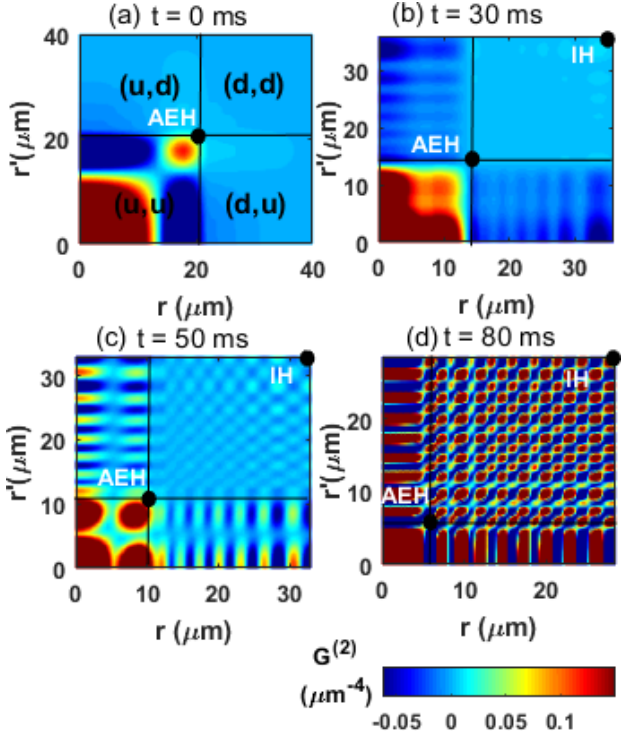}
\caption{{\it (color online).} $G^{(2)}(r,r';0^\circ)$ for $\eta'/\eta$ =  0.4 at various times, evaluated using consideration of atom number fluctuations in the simulation. AEH and IH are marked. }\label{numberf1}
\end{figure}

We also evaluate the correlation function by considering atom number fluctuation in the initial state, i.e. the ground state solution of the time-independent GPE (solved only in the presence of confining trap potential), $\psi_{\kappa 0}({\bf r})$. The variation in the number of atoms can always occur in BEC experiments. We have considered mean number of atoms in the condensate, $\bar{N}$ $\sim$ 6000 and standard deviation, $\Delta$N=0.15$\bar{N}$. The correlation function for $\zeta = 0^{\circ}$ at various times is shown in Fig. \ref{numberf1}. One of the dominant features observed in the supersonic region, similar to the previous TWA calculation, is the presence of checkerboard-like pattern formed at later times. But, here we do not observe one of the salient features from the correlation function, i.e. the signature of the presence of spontaneous HR as was seen in Fig.\ref{TWAzeta0}(b). We do not observe any off-diagonal dark bands in the upper-left quadrant (and lower right) originating from AEH which were earlier present in TWA calculation. We emphasise that correlation function measurements should be taken at controlled atom numbers as their variation, affects the correlation function in a quite similar manner to the TWA evaluation, especially at later times. 

The Fourier transform of individual quadrants of the correlation function gives plenty of information \cite{entanglementJS}. We will use this knowledge to extract the correlation spectrum between the points outside and inside the SBH in the next section. We shall mainly utilise the marked (u,d) quadrant from the density-density correlation function, obtained using TWA simulation, analysing the emitted radiation further.

\section{Spectrum of radiation}\label{SOR}
Analogue of Hawking pairs in SBH correspond to a pair of particles (phonons) created at the horizon simultaneously that propagate with equal and opposite energy, where one particle moves inside the SBH and the other outside it. In this section, we will utilise the correlation function evaluated using TWA, discussed in the previous section, to first identify the region in the correlation plot which corresponds to the correlations between the pair of distant points (r,r$'$) located on the opposite sides of the horizon with equal propagation times. For this purpose, we can utilise the top-left and the bottom-right quadrants of the correlation function as they contain the information about the correlation between the points located on opposite sides of the horizon. And later, we will use it to evaluate the spectrum of the radiation emitted from this SBH at few illustrative times in the simulation. 

\begin{figure}
\centering
\includegraphics[scale = 0.65]{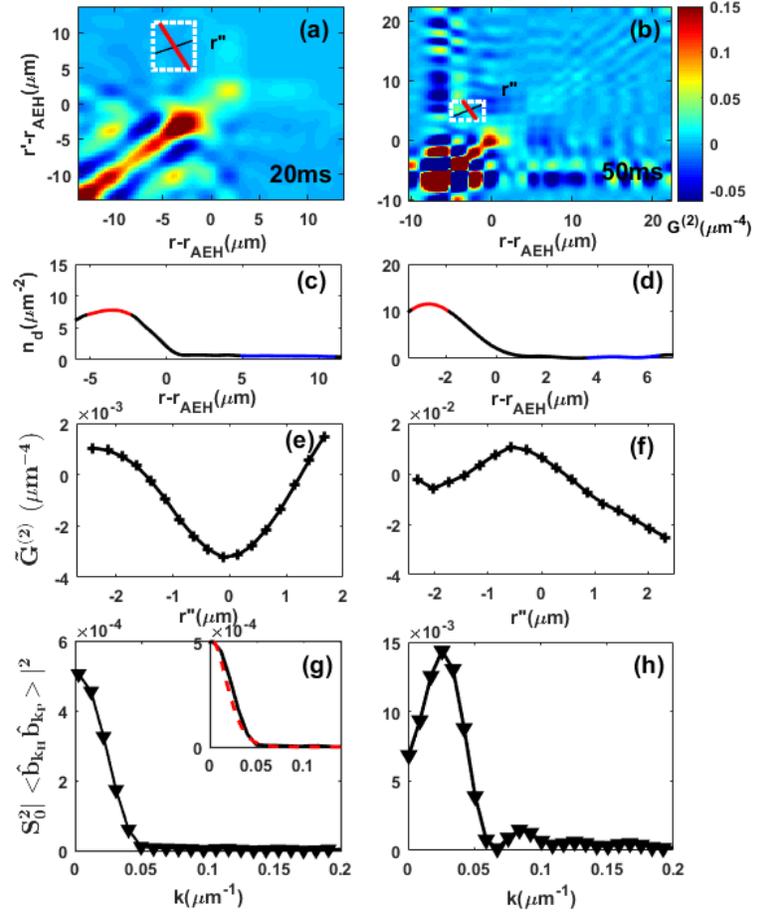}\\
\caption{{\it (color online).}(a-b) Correlation function (with re-scaled spatial coordinate) for $\eta'/\eta = 0.4$ along $\zeta =0^{\circ}$ at 20 and 50 ms respectively. Here, the red (thick) line corresponds to the correlation between the distant points located on opposite sides of the horizon with equal propagation times. The corresponding region consisting of these points are also marked in the total density (blue and red) in (c-d), respectively. (e-f) displays the profile of $G^{(2)}$ along the black line in (a-b), respectively. (g-h) shows the spectrum of the radiation corresponding to (a-b) at the respective times.}\label{thermala}
\end{figure}

To compute the spectrum of the emitted radiation, we take some of the results from Fig. \ref{TWAzeta0}. Further to simplify our analysis, we re-scale the absolute value of the radial coordinate r (and r$'$) to mark the location of AEH as the origin, by replacing it with $r-r_{AEH}$ ($r'-r_{AEH}$) in the correlation function where, as an example, $r_{AEH}$ =  (16.91, 10.28) $\mu$m at (20, 50) ms respectively, for the results depicted in Fig. \ref{TWAzeta0}(b, d) respectively. The correlation function after the re-scaling of the coordinate is shown, in Fig. \ref{thermala}(a-b) respectively. 

With the help of Fig. \ref{thermala}(a-b), we will first identify the region in one of the quadrants (top-left) that contains the information of the correlation between the points with equal propagation times. For this, we first note that the phonons simultaneously created at horizon propagate with speeds $c^{av}_{out}-v^{av}_{out}$ and $v^{av}_{in}-c^{av}_{in}$ outside and inside the BH respectively. At the time of emission,
\bea 
 r&=&(v^{av}_{out}-c^{av}_{out})\tau <0 \text{ in the outside region and, }\nn\\
r'&=&(v^{av}_{in}-c^{av}_{in})\tau >0 \text{ in the inside region of SBH}\nn
\eea
as $r,r'$ = 0 now corresponds to the AEH location after re-scaling of coordinate. Here, $\tau$ is the time of emission of Hawking phonons at the horizon, and $c^{av}_{in/out}, v^{av}_{in/out}$ are respectively the average sound and flow speeds in the inside(in) and the outside(out) region of the SBH. 

\begin{figure*}
\subfloat{\includegraphics[scale = 0.7]{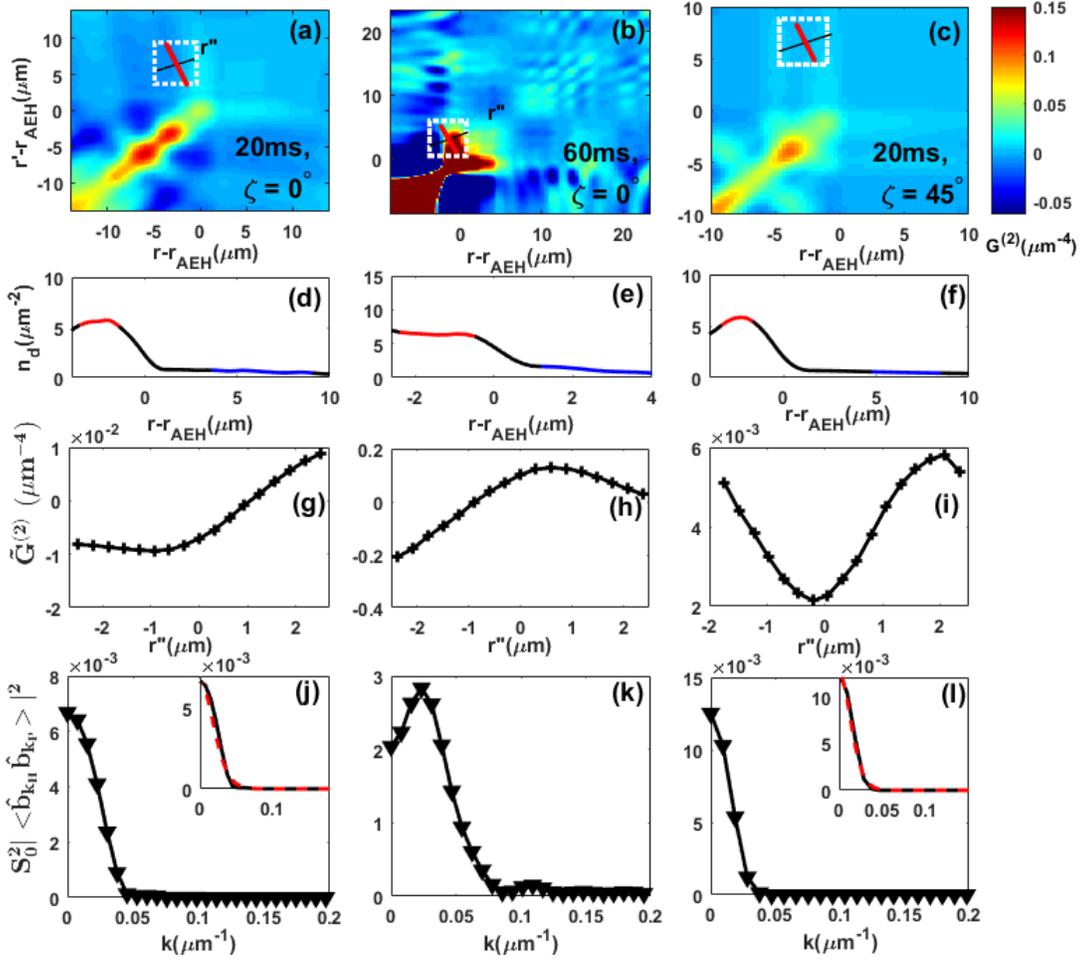}}
\caption{{\it (color online).}(a-c) Correlation function (with re-scaled spatial coordinate) for $\eta'/\eta = 0.78$ at various times mentioned and angle, $\zeta$ in each of the figure. (d-f) shows the corresponding density of outside (red) and inside (blue) regions of SBH, where the correlations are observed in (a-c) respectively. (g-i) displays the profile of $G^{(2)}$ along the black line in (a-c), respectively. (j-l) shows the spectrum of the radiation  emitted corresponding to (a-c) respectively at the mentioned times.}\label{thermalc}
\end{figure*}

The values of data corresponding to 20 ms in this case are: ($c^{av}_{in},c^{av}_{out}, v^{av}_{in}, v^{av}_{out})$ = (0.09, 0.30, 0.58, 0.07) mm/s and the respective values for 50 ms are : (0.09, 0.43, 0.65, 0.12) mm/s. Using these values, the line corresponding to the correlations between the pair of points ($r,r'$) located on opposite sides with equal propagation times, in the density-density correlation function, are identified by evaluating the slope through: 
\bea \frac{r'}{r} = \frac{v^{av}_{in}-c^{av}_{in}}{v^{av}_{out}-c^{av}_{out}} \label{slopeHR}
\eea
and, are marked in red (thick) in Fig. \ref{thermala} (a-b) for (20, 50) ms respectively.
Fig.\ref{thermala}(c-d) shows the regions in the total density, along $\zeta = 0^{\circ}$, where these correlations are observed. Blue region corresponds to the inside of SBH and red to the outside region. For $\eta'/\eta = 0.78$ the corresponding plots, are shown in Fig. \ref{thermalc}. The values of ($c^{av}_{in},c^{av}_{out}, v^{av}_{in}, v^{av}_{out})$ used for this case are (0.07, 0.29, 0.65, 0.07) mm/s and (0.12, 0.42, 0.88, 0.06) mm/s at (20, 60) ms respectively. We have also shown the corresponding plots for the initial time (20 ms) along $\zeta = 45^{\circ}$, for which the values are (0.04,0.30,0.65,0.05) mm/s. 

Earlier we have identified the bands originating from AEH, as the spontaneous Hawking radiation emitted at the initial time. As an example, consider the case at 20 ms for $\eta'/\eta = 0.4, \zeta = 0^\circ$ [Fig. \ref{TWAzeta0}(b)] in the previous section. And, here also this identification is made by marking the correlations between the pair of points ($r,r'$) for equal propagation times using Eq. (\ref{slopeHR}) for the corresponding case, in Fig.\ref{thermala}(a) [red (thick) line]. We will further extract information about the spectrum of the emitted radiation from this SBH at few of the illustrative times for both $\eta'/\eta = 0.4, 0.78$. We will focus on the (u,d) quadrant of the density-density correlation, namely, we will utilise the dashed rectangle marked in Fig.\ref{thermala}(a-b). The Fourier transform of the correlations in such a region gives the spectrum of correlations between the Hawking and partner modes, $<{{\hat{b}}_{k_{H}}}{{\hat{b}}_{k_{P}}} >$ \cite{entanglementJS} related explicitly as (Appendix \ref{TSHR}),
\bea
S_0 <{{\hat{b}}_{k_{H}}}{{\hat{b}}_{k_{P}}} >&=& \frac{1}{\sqrt{N^u N^d}}\int dr dr' e^{ik_H r}e^{ik_P r'} G^{(2)}(r,r')\label{radspec}\nn\\
\eea
where $\hat{b}_{k_{H}},\hat{b}_{k_{P}}$ are the annihilation operators for the Hawking mode (localised outside the SBH with wavenumber $k_H$) and partner mode (localised inside the SBH with wavenumber $k_P$) respectively; $N^u, N^d$ is the total number of atoms in the upstream and downstream regions respectively. $S_0$ is the static structure factor at zero temperature, $S_0 = ( u_{k_{H}} + v_{k_{H}})( u_{k_{P}} + v_{k_{P}})$, and u's and v's are the Bogoliubov coefficients for the phonons corresponding to the total density mode(Appendix \ref{TSHR}).

The Eq.(\ref{radspec}) is derived, by neglecting the correlations between phonons moving in opposite directions from the horizon with different frequencies\cite{Steinhauer16,entanglementJS} (refer Appendix \ref{TSHR}). We compute the integral in the RHS of Eq.(\ref{radspec}) using the method adapted in \cite{Steinhauer18,Pavloff20}, which comprises averaging $G^{(2)}$ in the region, inside
the rectangle marked in Fig.\ref{thermala}(a-b). To this purpose, a
local coordinate $r''$, orthogonal to the locus of points
along the red (thick) line marked in $G^{(2)}$ [Fig.\ref{thermala}(a-b)] is defined and then average is computed. The profile of $G^{(2)}$ along the black line (perpendicular to the red (thick) line), $\tilde{G}^{(2)}$ as a function of the variable $r''$, corresponding to Fig.\ref{thermala}(a-b), is shown in Fig.\ref{thermala}(e-f) respectively. For evaluating the spectrum, the Fourier transform of the average of $G^{(2)}$ profile is then computed, after zero-padding at its end \cite{zero} to yield larger closely-spaced frequency bins in the frequency domain. The spectrum $S_0^2|<{{\hat{b}}_{k_{H}}}{{\hat{b}}_{k_{P}}} >|^2$ \cite{Steinhauer18}, for the illustrated times, are plotted respectively in Fig. \ref{thermala}(g-h). The corresponding plots for $\eta'/\eta = 0.78$ along $\zeta = 0^\circ, 45^\circ$ at some of the illustrative times, are shown in Fig. \ref{thermalc}.


\begin{figure}
\centering 
\includegraphics[scale = 0.69]{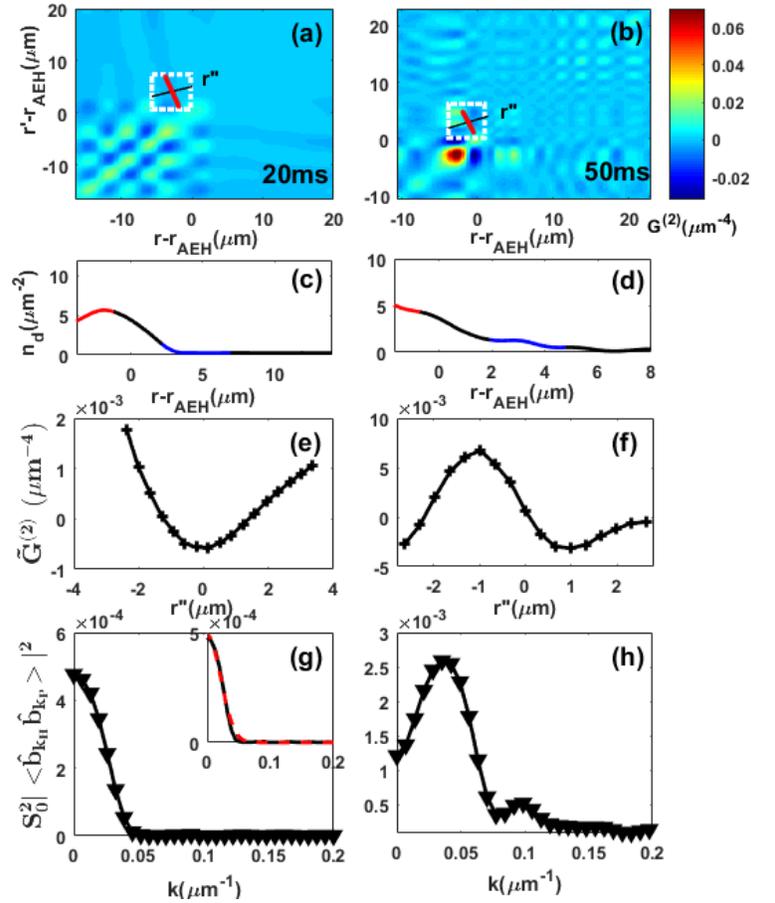}
\caption{{\it (color online).}(a-b) Correlation function (with re-scaled spatial coordinate) for a two-component BEC (without spin-orbit coupling), along $\zeta =0^{\circ}$, at 20 and 50 ms respectively. The red (thick) line corresponds to the correlation between the pair of distant points ($r,r'$) located on the opposite sides of the horizon, with equal propagation times. The corresponding region consisting of these points are also marked in the total density (blue and red) in (c-d), respectively. (e-f) displays the profile of $G^{(2)}$ along the black line in (a-b), respectively. (g-h) shows the spectrum of the radiation corresponding to (a-b) at the respective times. Inset shows the agreement between the spectrum evaluated using Eq.(\ref{radspec}) (solid curve) with that of the predicted thermal spectrum (dashed curve).
}\label{thermald}
\end{figure}
\begin{figure}
\centering 
\includegraphics[scale = 0.8]{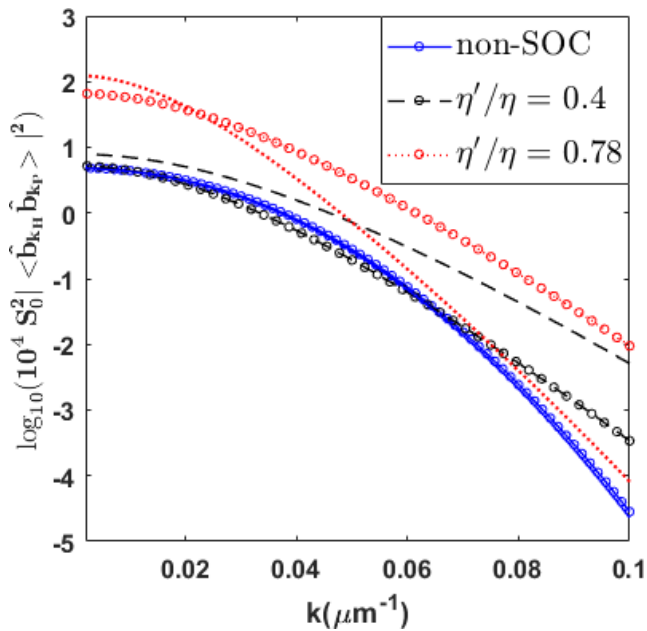}
\caption{{\it (color online).} Comparison of the predicted spectrum at 20 ms is shown, for the case of two-component BEC (without SOC) (solid curve blue) with two sets of SOC-strengths: $\eta'/\eta$ = 0.4 (dashed black) and 0.78 (dotted red). The results along $\zeta = 0^{\circ}$ are labelled, with a marker `o' and, results along $\zeta = 45^{\circ}$ are without any marker. }\label{thermale}
\end{figure}
In the above discussion, we have numerically computed the spectrum of radiation emitted from the SBH, utilising the density-density correlation function. We can also compute the predicted spectrum of the analogue Hawking radiation emitted from the SBH at the initial times, by evaluating the expression ``$S_0^2 |\beta|^2 (1+|\beta|^2)$", where we have used $|<{{\hat{b}}_{k_{H}}}{{\hat{b}}_{k_{P}}} >|^2  = |\beta|^2 (1+|\beta|^2)$ \cite{Steinhauer18}. However, the utilisation of this expression is strictly valid in the liner-dispersion regime that corresponds to the initial times taken in our simulation, where the sonic analogy is valid. Here, $\beta$ is the Hawking parameter whose squared modulus follows the distribution, $|\beta|^2 = [\exp(\frac{\hbar \Omega_d}{k_B T_H})-1]^{-1}$. Thus to compute the predicted spectrum, we will require the knowledge of the dispersion, $\Omega_d$ in the subsonic region (Appendix \ref{2}) and predicted analogue Hawking temperature $T_H$ at AEH (Appendix \ref{TSHR}).  Since, the calculation of the $T_{H}$ (using the gravitational analogy) in the presence of SOC is different as compared to the one, given in the literature, for the case of no spin-orbit coupling (scalar condensate), we have provided the details in Appendix \ref{TSHR}. 

The predicted thermal spectrum, marked with the red dashed line, is shown in the inset of Fig. \ref{thermala}(g), \ref{thermalc}(j,l)  at one of the initial times (20 ms). The spectrum of radiation evaluated using Eq.(\ref{HT}) and Eq.(\ref{radspec}), for this case, are both in good agreement. Thus, we can conclude that the spectrum of the analogue Hawking radiation emitted at the initial times from this SBH is thermal. However, for later times illustrated in this work, the spectrum of emitted radiation is no longer thermal as shown in \ref{thermala}(h) and \ref{thermalc}(k) apparently, since it is not proportional to $\sim$ ``1/k" \cite{Steinhauer16} for smaller values of wavenumber, k and also have small bumps at higher k values. Such bumps, in the spectra, were also seen in \cite{Steinhauer16,Steintheory}. Thus, studying the long-term behaviour of this SBH configuration at later times, where the sonic analogy is no longer valid, illustrates that thermal radiation cannot be steadily emitted, due to the mechanism of black hole lasing, exhibited by this system.

The observation of thermal Hawking radiation at initial times and departure from thermality at later times are also being observed, in the case of a two-component BEC without spin-orbit coupling (see the end of Appendix \ref{2b}). The computation details involved for evaluating the spectrum of such a non-SOC configuration are shown, in Fig. \ref{thermald}. In this figure, the spectrum of the emitted radiation at representative times 20 ms and 50 ms along $\zeta = 0^\circ$ are plotted, in (g-h) respectively. Additionally, in Appendix \ref{2b},
the density-density correlation function at various times is also shown in Fig. \ref{tc_corr}. The values of ($c^{av}_{in}, c^{av}_{out}, v^{av}_{in}, v^{av}_{out})$ used to mark the correlations between the pair of points ($r,r'$) with equal propagation times (red line) in Fig. \ref{thermald} (a-b), for these illustrative times are: (0.07, 0.28, 0.55, 0.12) mm/s and (0.10, 0.42, 0.60, 0.14) mm/s respectively.

Our comparison of the predicted spectrum of analogue HR emitted from the SBH realised from SOC-BEC with that of the non-SOC case discussed above, at an initial time (20 ms), in Fig. \ref{thermale} shows the strong impact of SOC on such analogue HR. 
For SOC configuration, the spectrum of emitted radiation at a particular time, along the different azimuthal direction, varies regarding its spectral properties such as height and width, as illustrated in Fig. \ref{thermalc} (j,l). The difference among various directions is significantly larger for a higher ratio of $\eta'/\eta=0.78$, shown for $\zeta=0^{\circ}$ and $45^{\circ}$, in Fig. \ref{thermale}. Whereas, the non-SOC configuration (two-component BEC) exhibits isotropic spectral properties among these directions, at a given time (refer Fig. \ref{thermale}). Quantitatively, the height of the spectrum for a higher ratio of SOC strengths differs almost by order of magnitude from that of non-SOC case. However, the spectrum's height in the non-SOC case is of the same order as that of SOC case with a smaller ratio of SOC strengths (i.e. $\eta'/\eta$ = 0.4). It may also be noted that the expression for the analogue Hawking temperature [Eq. (\ref{HT})], calculated using the gravitational analogy also demonstrates the additional contribution due to SOC to the $T_H$ (see the last part of Appendix \ref{TSHR}). 
\section{Conclusion}
In conclusion, we showed that a SOC-BEC in a suitable laser-driven potential, in the absence of any external rotation, can realise an analogue SBH which emits analogue Hawking radiation. The break down of the irrotationality condition in such SOC-BEC, even without any externally imposed rotation is marked, by a considerable azimuthal flow at a comparatively larger value of the anisotropy parameter. The space-time metric of such SBH formed out of SOC-BEC, under some limiting condition, can be mapped to that of a $2+1$ dimensional black Hole \cite{BTZ, BTZrev} within the hydrodynamic approximation. 
The azimuthal flow, for the parameters studied in this work, is however incoherent and improvement of this aspect requires further studies in future.  

 Integrating the GPE by considering quantum fluctuations in the initial state using TWA in the presence of laser-driven potential for a substantial period, we have also analysed the analogue Hawking radiation emitted from this SBH and its corresponding radiation spectrum. At the initial times, in the simulation, observation of negative correlations between the points lying on the opposite sides of the horizon for equal propagation times is seen very clearly. Hence spontaneous Hawking radiation is identified, in the considered system. Also, a deviation from the thermal behaviour of the analogue Hawking radiation is seen, at later times in the simulation due to the black hole laser effect. We also compare the results with a non-SOC case and show how the spin-orbit coupling impacts on the sonic Hawking radiation. We hope our work will lead to further theoretical and experimental studies on the various models of SOC-BEC as a possible analogue model of rotating SBH in different dimensions. 

\begin{center}
{\bf ACKNOWLEDGEMENTS}
\end{center}
We thank J. Steinhauer for critical reading of the earlier version of this manuscript and helpful comments. SG also thanks I. B. Spielman and I. Carusotto for helpful discussion.
The work is supported by a BRNS (DAE, Govt. of India) Grant no. 21/07/2015-BRNS/35041 (DAE SRC  Outstanding Investigator scheme). IK is supported by a fellowship by MHRD, Govt. of India.


\appendix
\section{Derivation of the GP Equation from the microscopic hamiltonian}\label{A}
In this section, we will provide details of the derivation of GPE [Eq.(\ref{gpe})] from the microscopic Hamiltonian that closely follows refs. \cite{system1,brandon}. Section \ref{1a} starts with the physical description of the system considered and its single-particle and many-body Hamiltonian. In section \ref{1b}, we provide the details of projection in the lower energy subspace and the corresponding GPE for this system. A brief summary of (\cite{system1,brandon}) is provided, for the sake of completeness, in the following part.

\subsection{The System and the microscopic Hamiltonian}\label{1a}

\begin{figure}[h!]
\includegraphics[scale=0.48]{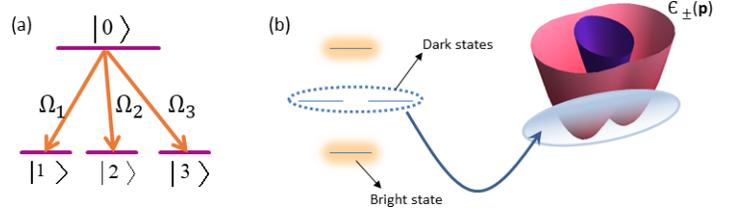}
\centering
\caption{{\it (color online).} (a) Shows the basic tripod system, (b) shows the four states obtained from the atom-laser interaction Hamiltonian and a schematic that we are going from dark states  to lower energy subspace.}\label{tripod}
\end{figure}

 The spin-orbit (SO) coupling we consider in this work, can be generated using a tripod scheme in which three degenerate atomic states $\ket{1},\ket{2},\ket{3}$ are coupled to an excited state $\ket{0}$ (Fig.\ref{tripod}(a)) via three optical fields that generate an effective non-abelian gauge potential equivalent to SO interaction \cite{system1,brandon}. The single-particle Hamiltonian of an atom-laser interacting system is given as, $H=(\frac{P^2}{2m}+V)\check{I}+H_{al}$ with the atom-laser interaction Hamiltonian, 
$ H_{al}=\Omega_0\ket{0}\bra{0}+\bigg[\Omega_1\ket{0}\bra{1}+\Omega_2\ket{0}\bra{2}+\Omega_3\ket{0}\bra{3}+h.c\bigg]$
where Rabi frequencies $\Omega_\mu$ ($\mu=0,1,2,3$) are parametrised with space and phase variables.

 This scheme gives two degenerate dark states and two bright states. The dark states are weakly coupled to the other two states and hence, adiabatically eliminating bright states gives us the following Schrodinger equation \cite{system1} :

\beq  i\hbar \frac{\partial }{\partial t}  \psi  = \bigg[ \frac{1}{2m}
   (-i\hbar\nabla-{\bf A})^2 + V+{W} \bigg] \psi  \nn \eeq 
where $
{\bf A}=m( \eta \hat{x} \check{\sigma}_y +  \eta' \hat{y}\check{\sigma}_z)$ is the non-abelian gauge potential,
${W}
=({A}^2)_{ii}-{A}_{ii}\cdot {A}_{ii}$ is the scalar potential which we shall ignore in this work, and $\psi = [\psi_{\uparrow}, \psi_{\downarrow}]^T$ refers to the wavefunction of two pseudo-spin components corresponding to degenerate dark states.
To study the BEC formed with such bosons we begin with the many-body Hamiltonian written in second quantised form \cite{brandon}: 
 \\
\bea
&\hat{H}&=\sum_{\gamma,\beta}\bigg[\int d{\bf r} \hat{\psi}^\dagger_{\gamma}({\bf r})(\hat{h}_{\gamma \beta}+V) \hat{\psi}_{\beta}({\bf r})\nn\\
&+&\frac{1}{2}\int d{\bf r}\int d{\bf r'} V_{int}({\bf r,r'})\hat{\psi}^\dagger_{\gamma} ({\bf r}) \hat{\psi}_{\beta} ({\bf r'}) \hat{\psi}^\dagger_{\beta} ({\bf r'})  \hat{\psi}_{\gamma} ({\bf r} ) \bigg]\nn\\\label{sqH}
\eea
where, the single particle Hamiltonian $\hat{h}_{\gamma \beta}=\bigg\{\frac{{\bf p}^2}{2 m}\check{I}-\eta {p}_x \check{\sigma}_y-\eta'{p}_y \check{\sigma}_z\bigg\}_{\gamma \beta}
$.
Here, we have considered the density-density interaction potential as a contact pseudopotential, $V_{int}({\bf r,r'})\sim V_{int} \delta({\bf r}-{\bf r'})$; $\hat{\psi}_{\gamma} ({\bf r} )$ are the field operators for bosons in pseudo-spin states, $\gamma=\uparrow,\downarrow$ and  V is the external potential. \\

\subsection{Projection to lower energy subspace and deriving GP Eq.(\ref{gpe})}\label{1b}
A suitable choice for the basis states can be constructed, by diagonalising the single-particle Hamiltonian which yields dispersion relation $\epsilon_{\lambda}({\bf p })= \frac{{\bf p }^2}{2m}+\lambda \sqrt{\eta^2 p_x^2 +\eta'^2 p_y^2}
$ where $\lambda=\pm1$ labels the bands.  In the further description, we shall be focussing only on the lower energy band corresponding to $\lambda=-1$. For isotropic SOC strengths, there's infinite degeneracy, and we have a Rashba ring; for anisotropic SOC strengths, $(\eta\neq \eta')$ there are two minima in the lower energy band which can be  in either x or y direction (in momentum space) depending on the magnitude of SOC strengths, $\eta$ and $\eta'$. Here in this work, we shall only consider the case where minima's occur at $\pm m\eta \hat{x}$, i.e. for the case, $\eta'< \eta$. The spectrum near
minima of the band corresponding to $\lambda=-1$, for small momentum fluctuations `{\bf p}' around the minima $\pm m\eta \hat{x}$, i.e. replacing ${\bf p} \rightarrow \pm m\eta \hat{x}+{\bf p}$ with $|{\bf p}|<<m\eta$, we get the effective dispersion as:
 $E_-({\bf p})=\frac{p_x^2}{2m} + \frac{p_y^2}{2m_y}+\frac{p_z^2}{2m}$, where $m_y$ is the effective mass along y-direction that represents the curvature in the minima. A single-band model is sufficient to describe the BEC physics \cite{negativemass,spindensity}. Therefore, we focus on the lower band (-) and restrict the range of {\bf p} near to right $({\bf p_{+}}=m\eta \hat{x})$ and left $({\bf p_{-}}=-m\eta\hat{x})$ minima. The components of the eigenfunction corresponding to the lower band, is given, as ${\psi^\gamma}_{ {\bf p}-}({\bf r})=e^{i{\bf p \cdot r}}U_{-1\gamma}(\chi_{\bf p})$, $\text{where} \hspace{0.1cm}\chi_{\bf p}=\tan^{-1}(p_y/p_x)$, with
\begin{equation}
U_{-1\uparrow}=\frac{[
\sqrt{\cos^2 \chi_{\bf p} +\Delta^2\sin^2 \chi_{\bf p}}+\Delta\sin \chi_{\bf p}]^{1/2}}{\sqrt{2}[\cos^2 \chi_{\bf p} +\Delta^2\sin^2 \chi_{\bf p}]^{1/4}} \nn
\end{equation} 
\begin{equation}
U_{-1\downarrow}= i \textit{sgn}[\cos \chi_{\bf p}] \frac{[
\sqrt{\cos^2 \chi_{\bf p} +\Delta^2\sin^2 \chi_{\bf p}}-\Delta\sin \chi_{\bf p}]^{1/2}}{\sqrt{2}[\cos^2 \chi_{\bf p} +\Delta^2\sin^2 \chi_{\bf p}]^{1/4}}\nn
\end{equation}
where $\Delta = \eta'/\eta$.
Using these, the field operators in Eq.(\ref{sqH}) are expanded, in terms of the states near the minima as, $\hat{\psi}_{\gamma}({\bf r})={\sum_{{\bf p},\kappa}} \hat{\psi}_-({ {\bf p}+{\bf p_{\kappa}}}){\psi^\gamma}_{ {\bf p}+{\bf p_{\kappa}}-}({\bf r}) $. For notational convenience, we are replacing $\hat{B}_{\kappa({\bf p+p_\kappa})}$ with $\hat{\psi}_-({ {\bf p}+{\bf p_{\kappa}}})$. Evaluating the commutator $[\hat{\psi}_{\gamma}({\bf r}),\hat{H}]$ 
 and replacing operator $\hat{\psi'}_\kappa({\bf p})$ with the order parameter ${\psi'}_\kappa({\bf p})$, in the mean-field approximation \cite{pitaevski}, one yields the following multicomponent Gross-Pitaevskii equation: 
\bea 
 i\hbar\frac{\partial \psi'_{\kappa} }{\partial t}&=&\bigg[\frac{\hbar^2}{2m}(-i\partial_x-\kappa \frac{m \eta}{\hbar})^2-\frac{\hbar^2}{2m_y}\partial_y^2-\frac{\hbar^2}{2m}\partial_z^2\bigg]\psi'_{\kappa}\nn\\
  &+& V_{int}\bigg( |\psi'_-|^2+|\psi'_+|^2\bigg)\psi'_{\kappa} \nn \eea
where we have written, $\hat{\psi}_-({ {\bf p}+{\bf p_{\kappa}}})=\hat{\psi'}_\kappa({\bf p})$ and used $\hat{\psi'}_\kappa({\bf r})=\sum_{{\bf p}}e^{i{\bf p \cdot r}}\hat{\psi'}_\kappa({\bf p})$ (for more details, refer \cite{brandon,Quinlu}). To study the properties of the analogue sonic black hole in such system, we study the dynamics of such pseudospin-$\frac{1}{2}$ bosons in a time-dependent potential, $V=V_{Trap}+V_{step}$,
where $V_{Trap}$ is the harmonic confinement and $V_{step}$ is the step potential defined in the main text. 
In our simulations, we considered the trap frequencies $\omega_{x,y,z}=2\pi \times \{4.5,4.5,123\}$ Hz. Thus, the strong confinement along the z-direction reduces the dimension to quasi-2D \cite{Wbao,parola} and hence, we obtain the 2D-GPE, Eq.(\ref{gpe}).


\section{Bogoliubov dispersion of SOC BEC and the sound velocities}\label{2}
Following \cite{brandon}, we first discuss the Bogoliubov dispersion for the condensate of such SOC bosons in this section and then calculate the sound velocities from this dispersion. 

The Hamiltonian corresponding to Eq.(\ref{sqH}), in momentum space, is given as:
\begin{eqnarray}
H 
&=&  \sum_{\gamma,\beta;{\bf p}} \hat{b}^\dagger_{\gamma {\bf p}}   \hat{h}_{\gamma\beta} \hat{b}_{\beta {\bf p}} +\frac{ V_{int}}{2\mathcal{V}} \sum_{\gamma,\beta}  \sum_{{\bf p,p',q}}  \hat{b}^\dagger_{\alpha {\bf p}}\hat{b}_{\gamma {\bf p}+{\bf q}}  \hat{b}^\dagger_{\beta {\bf p'}}   \hat{b}_{\beta {\bf p'}-{\bf q}}\hspace{0.1cm} \text{,}  \nn
\end{eqnarray}
\text{ where } $\mathcal{V}=(2 \pi)^3$ and $\hat{b}_{\gamma {\bf p}}$ is the annihilation operator in the dark-state subspace with momentum {\bf p} and pseudo-spin $\gamma$ related to the field operators in Eq.(\ref{sqH}) through Fourier transform.
The Hamiltonian is then projected onto the lower band through the operators, $\hat{B}^\dagger_{\lambda {\bf p}}= \hat{b}^\dagger_{\gamma {\bf p}} {U}_{\gamma\lambda}({\bf p})$ and then described for a sector (n, N-n) left and right movers, in terms of left/right well (L/R) operators as $\hat{B}_{L/R {\bf p}}=\hat{B}_{-1\mp ({\bf p}+m\eta)}$, $U_{L/R \gamma}({\bf p})=U_{-1\gamma}{(\mp [{\bf p}+m\eta])}$ (for details refer \cite{brandon}), given by:
\begin{eqnarray}
H_{int}&=&\frac{V_{int}}{2\mathcal{V}}\sum_{{\bf p,p',q}} {\sum_{\{\sigma_i=L/R\}}}'\hat{B}^\dagger_{\sigma_1{\bf p}}\hat{B}_{\sigma_2{\bf p}+{\bf q}}\hat{B}^\dagger_{\sigma_3{\bf p'}}\hat{B}_{\sigma_4{\bf p'}-{\bf q}}\times\nn\\
&U&^\dagger_{\sigma_1\gamma}({\bf p}) U_{\sigma_2 \gamma}({\bf p}+{\bf q}) U^\dagger_{\sigma_3 \beta}({\bf p'})   U_{\sigma_4\beta}({\bf p'}-{\bf q})\label{Hlowersubspace}
\end{eqnarray}
where $\sum'$ indicates that the sum is restricted to the equal number of left and right movers.
In order to diagonalise the above hamiltonian, the following Bosonic operators, are introduced
\[
 \begin{bmatrix}
    \hat{B}_{-,{\bf p}} \\ \hat{B}_{+,{\bf p}} \\
  \end{bmatrix}\\ =
  \begin{bmatrix}
   \sqrt{1-\frac{n}{N}} e^{-i\chi/2}&  -\sqrt{\frac{n}{N}}e^{i\chi/2} \\
     \sqrt{\frac{n}{N}}e^{-i\chi/2} &  \sqrt{1-\frac{n}{N}} e^{i\chi/2}\\
  \end{bmatrix}
  \begin{bmatrix}
    \hat{B}_{L{\bf p}} \\ \hat{B}_{R{\bf p}} \\
  \end{bmatrix}, \\
  \]
where $\chi$ is some arbitrary phase.
This transformation makes the Hamiltonian in Eq.(\ref{Hlowersubspace}) partially diagonal. It can be viewed as the as a symmetric and antisymmetric combination of the two basis state operators $(\hat{B}_{L{\bf p}}$ and $\hat{B}_{R{\bf p}})$, as done usually for double-well case. Subsequently introducing $\hat{\beta}_{-,{\bf p}}=\hat{B}_{-,{\bf p}}$
 and $\hat{\beta}_{+,{\bf p}}=(1-A_{\bf p}^2)^{-1/2}(\hat{B}_{+,{\bf p}}-A_{\bf p} \hat{B}^\dagger_{+,-{\bf p}})$, where $
A_{\bf p}=\frac{1}{2}[-(s({\bf p})+2)+ \sqrt{s({\bf p})(s({\bf p})+4)}]$ with $s({\bf p})=2 E_-({\bf p})/[nV_{int}({\bf p})]$, makes the effective low-energy many body Hamiltonian completely diagonal,
  \begin{eqnarray}
\mathcal{H}&=&E^{(0)}_0+\hbar \sum_{{\bf p}}[\Omega_z({\bf p})\hat{\beta}^\dagger_{-,{\bf p}}\hat{\beta}_{-,{\bf p}}+ \Omega_d({\bf p})\hat{\beta}^\dagger_{+,{\bf p}}\hat{\beta}_{+,{\bf p}}]\nn
\end{eqnarray}
where first term is the condensate energy and $
 \Omega_{d,z}({\bf p})$ represents the quasiparticle excitation spectrum for the total density and polarization density modes respectively.
\bea
 \hbar\Omega_d &=& \sqrt{E_{-}(\bs{p})[E_{-}(\bs{p}) +2 g \bar{n}_d]},\bar{n}_d=N/\mathcal{V}\nn\\
 \hbar\Omega_z &=& E_{-}(\bs{p})\nn
\eea
Total density modes correspond to the density of the total number of atoms and polarization density modes refers to the density corresponding to the relative number of atoms. 

Sound velocity for total density and polarization density modes are given by, 
 $c^{x,y}_{s}= \hbar \frac{d\Omega_d}{dp_{x,y} }\bigg|_{{\bf p}  \rightarrow 0}$ and $c^{x,y}_z = \hbar \frac{d\Omega_z}{dp_{x,y} } \bigg|_{{\bf p} \rightarrow 0}=0$ respectively, where ${\bf p}$ can be chosen either along x or y direction. Explicitly, the anisotropic sound velocities corresponding to total density modes are given, as $c^x= \hbar  \frac{d\Omega_d}{d{p_x} }\bigg|_{{p_x}  \rightarrow 0}=\sqrt{\frac{g \bar{{n}}_d}{m}}$, $c^y= \hbar \frac{d\Omega_d}{d{p_y} }\bigg|_{{p_y}  \rightarrow 0}=\sqrt{\frac{g \bar{{n}}_d}{m_y}}$. 
\begin{figure} 
\includegraphics[scale=0.6]{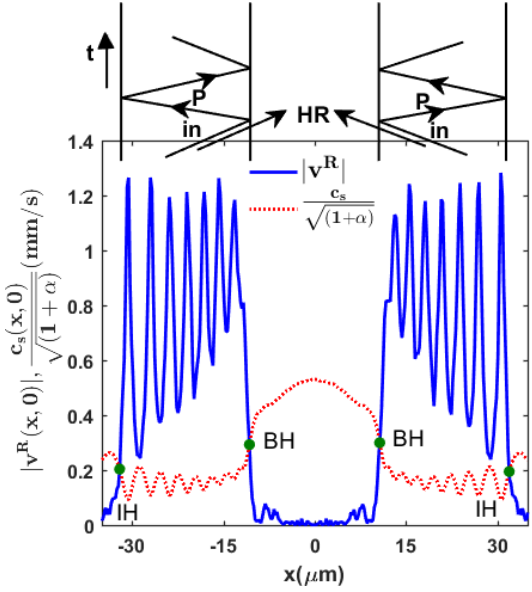}
\caption{{\it (color online).} We plot the one-dimensional cross-sectional view of the subsonic and supersonic regimes at 50 ms for $\eta'/\eta$=0.78 superposed with the  space-time diagram.}\label{Spacetime}
\end{figure} 
Bogoliubov dispersion for the quasi-two-dimensional SOC-BEC corresponding to the total density, $n_d=n_+ + n_-$ is given as: \bea
\Omega_d^2 &=& \frac{\hbar^2}{4m^2}\bigg[k_x^2 + \alpha k_y^2 \bigg]^2 + c_x^2 \bigg[k_x^2 + \alpha k_y^2 \bigg] \nn\\
&=& \frac{\hbar^2 K^4}{4m^2} +   K^2 \frac{c^2}{(1+\alpha)}\label{dispSOC}
\eea
where, $K^2 = k_x^2 + \alpha k_y^2$.
Eq.(\ref{dispSOC}) is the dispersion in the comoving frame of the condensate. Although a spin-orbit coupled BEC breaks Galilean invariance, at low energies, it obeys the Galilean invariance condition \cite{GI1,GI2}. To get the dispersion in the observer frame, we make the Galilean transformation, $\Omega_d \rightarrow \Omega_d - {\bf v \cdot K} $ and the local dispersion relation in the supersonic and subsonic regimes in this frame of reference are thus, obtained.
\begin{figure}
\includegraphics[scale=0.6]{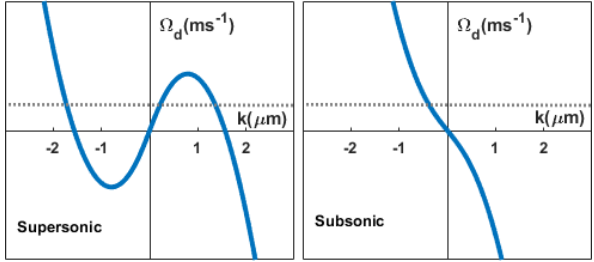}
\caption{{\it (color online).} Dispersion relation in subsonic region and supersonic regions for $\eta'/\eta = 0.4$ at 20 ms along $\zeta = 0^\circ$.}\label{sup_disp}
\end{figure}
\bea
(\Omega_d - {\bf v \cdot K} )^2 = \frac{\hbar^2 K^4}{4m^2} +   K^2 \frac{c^2}{(1+\alpha)}\label{dispSOC}
\eea
The formula depends on the angle between ${\bf v, K} $. Thus, for illustration, we show in Fig. \ref{sup_disp}, the dispersion in the subsonic and supersonic regimes corresponding to $\zeta = 0^{\circ}$, at one of the times. We have shown only the relevant branches here. At constant energy (marked) in the dispersion figure, there
are two solutions in the supersonic region of opposite energy, whereas, there is one in the subsonic region. The presence of these two solutions in the supersonic region makes the onset of Hawking physics possible in such analogue systems, i.e. pairs of particles with opposite
energy can be created while conserving the total energy of the system. 
\section{expression for current and Hydrodynamic description   }\label{2b}
 In this section, we will first derive the Eq.(\ref{vel}) of this work and then describe the Hydrodynamic formalism for the system considered.
 We begin with writing the equation in terms of density of individual components, obtained from Eq.(\ref{gpe}), given as:
 \bea 
 \frac{\partial n_{+}}{\partial t} &=& \partial_x \bigg[\frac{-{\hbar}}{2 i m}\bigg(  {\psi_{+}}^* \partial_x \psi_{+} - \psi_{+} \partial_x {\psi_{+}}^* \bigg)+ \eta  n_{+}\bigg]\nn\\
 &-&\frac{{\hbar}}{2 i m_y}\partial_y\bigg[ {\psi_{+}}^* \partial_y \psi_{+}- \psi_{+} \partial_y {\psi_{+}}^* \bigg]  \label{3a}\\
\frac{\partial n_{-}}{\partial t}&=&\partial_x \bigg[ \frac{-{\hbar}}{2 i m}  \bigg( {\psi_{-}}^* \partial_x \psi_{-} - \psi_{-} \partial_x {\psi_{-}}^* \bigg)-  \eta n_{-}\bigg]\nn\\
&-&\frac{{\hbar}}{2 i m_y}\partial_y\bigg[ {\psi_{-}}^* \partial_y \psi_{-}- \psi_{-} \partial_y {\psi_{-}}^* \bigg]  \label{3b} \eea 

The wavefunctions, $\psi_{\pm}$ are different from the wavefunction of two pseudo-spin components which correspond to dark state subspace, $\psi_{\uparrow,\downarrow}$ defined in section \ref{1a}, as these are the two-component order parameters obtained after projecting to lower energy subspace. Adding and subtracting Eq.(\ref{3a}-\ref{3b}) gives,
\bea
\frac{\partial n_{d}}{\partial t}+\nabla \cdot {\bf j}_d &=& 0,
\frac{\partial s_{z}}{\partial t}+\nabla \cdot {\bf j}_z = 0 \nn
\eea
where ${\bf j}_{d,z}$ is the current for total density and polarization density, respectively whose x,y components are given, as:
\bea
{j_{d,z}}^x &=& \frac{{\hbar}}{2 i m}\bigg(  {\psi_{+}}^* \partial_x \psi_{+} - \psi_{+} \partial_x {\psi_{+}}^* \bigg)\nn\\ &\pm& \frac{{\hbar}}{2 i m}\bigg(  {\psi_{-}}^* \partial_x \psi_{-} - \psi_{-} \partial_x {\psi_{-}}^* \bigg)-\eta  s_{z}\label{current_d} \\
{j_{d,z}}^y &=& \frac{{\hbar}}{2 i m_y}\bigg(  {\psi_{+}}^* \partial_y \psi_{+} - \psi_{+} \partial_y {\psi_{+}}^* \bigg) \nn\\ &\pm& \frac{{\hbar}}{2 i m_y}\bigg(  {\psi_{-}}^* \partial_y \psi_{-} - \psi_{-} \partial_y {\psi_{-}}^* \bigg)\eea

Rewriting Eq.(\ref{current_d}) for total density mode in a condensed form gives Eq.(\ref{vel}). 
A one-dimensional cross-sectional plot corresponding to the condition for getting an event horizon [Eq (\ref{ergo2})] is shown, in  Fig. \ref{Spacetime}. The inner horizons (IH) and the Acoustic Event Horizon (AEH) horizons are marked. In comparison to the recent works \cite{Steinhauer14,Steinhauer16}, we have considered a closed geometry, keeping in mind two-dimensional nature of the problem.

Substituting $\psi_\kappa=\sqrt{n_{\kappa}} e^{i\theta_\kappa}$ in Eq. (\ref{gpe}),
where $n_{\kappa}$ and $\theta_\kappa$ are the densities and phases of the two components, respectively we get:
\bea
\frac{\partial n_\kappa}{\partial t}&=&-\partial_x [ n_\kappa (\frac{\hbar}{m}\partial_x \theta_\kappa- \kappa \eta) ]-\partial_y [\frac{\hbar}{m_y} n_\kappa \partial_y \theta_\kappa ]\label{hydro1}\\
\hbar\frac{\partial \theta_\kappa}{\partial t}&=&\frac{\hbar^2}{2m}\bigg[\frac{\partial_x^2  \sqrt{n_\kappa}}{ \sqrt{n_\kappa}}
- (\partial_x \theta_\kappa)^2 \bigg]+
\frac{\hbar^2}{2m_y}\bigg[\frac{\partial_y^2  \sqrt{n_\kappa}}{ \sqrt{n_\kappa}} - (\partial_y \theta_\kappa)^2 \bigg]\nn\\ &-&\kappa(\frac{m\eta^2}{2}+V_{2D})
-g_{2D}(n_-+n_+)+\kappa \hbar \eta \partial_x \theta_\kappa 
\label{hydro2}
\eea

   \begin{figure*}
\includegraphics[scale=0.65]{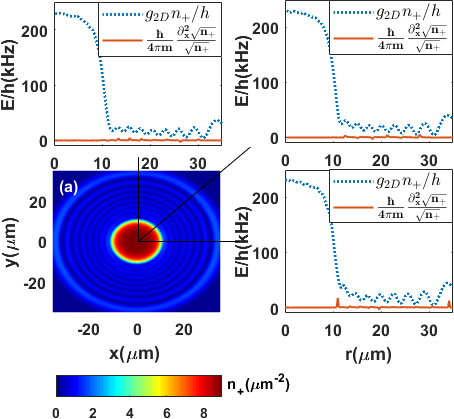}
\includegraphics[scale=0.65]{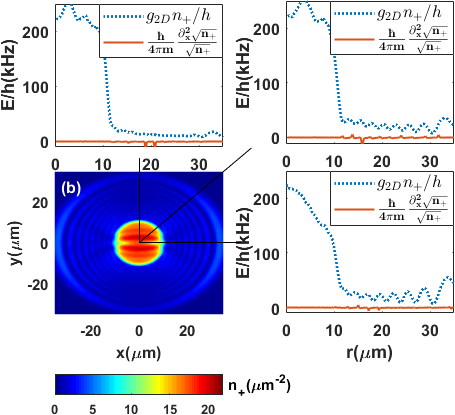}\centering
\caption{{\it (color online).} Comparision of quantum pressure term with the interaction energy is shown for the cross-section of density component, $n_+$ along $0^{\circ}, 45^{\circ}$ and 90$^{\circ}$ for $\eta'/\eta =$ (a) 0.4 (in left), (b) 0.78 (in right) at t=50 ms.}\label{hydroapprox}
\end{figure*}
Eq.(\ref{hydro1}) is the continuity equation satisfied by the two components. The quantum pressure term ($ \sim \frac{\partial_i^2  \sqrt{n_\kappa}}{ \sqrt{n_\kappa}} $; i=x,y) is very small as compared to the interaction energy, as shown in Fig. \ref{hydroapprox}, for one of the components. When the gradient in the condensate density is small, one can neglect the
quantum pressure term (from Eq.(\ref{hydro2})) leading to the hydrodynamic approximation \cite{visser11}.
Considering fluctuations ($\tilde{n}_{\kappa},\tilde{\theta}_{\kappa}$) around the mean value,
$n_\kappa\rightarrow \bar{{n}}_{\kappa}+ \tilde{n}_{\kappa}$, $
\theta_\kappa\rightarrow \bar{{\theta}}_{\kappa}+ \tilde{\theta}_{\kappa }$, where $\bar{n}_k = <n_k>$ and $\bar{\theta}_k = <\theta_k>$ are the background
density and phase of the components, respectively. Substituting in Eq.(\ref{hydro1}-\ref{hydro2}) and separating the background part, we get the following coupled differential equations:
\begin{eqnarray}
\frac{\partial \tilde{n}_{+ }}{\partial t}&=&-\partial_x[ \frac{\hbar  \bar{n}_{+}}{m}  \partial_x \tilde{\theta}_{+}+ \tilde{n}_{+} {v}_{+}^x]-\partial_y[ \frac{ \hbar \bar{n}_{+}}{m_y}  \partial_y \tilde{\theta}_{+}+ \tilde{n}_{+} {v}_{+}^y]\label{hydro7}\nn\\\\
\frac{\partial \tilde{n}_{-}}{\partial t}&=&-\partial_x[ \frac{ \hbar \bar{n}_{-}}{m}  \partial_x \tilde{\theta}_{-}+ \tilde{n}_{-} {v}_{-}^x]-\partial_y[ \frac{ \hbar \bar{n}_{-}\hbar}{m_y}  \partial_y \tilde{\theta}_{-}+ \tilde{n}_{-} {v}_{-}^y]\nn\\\\\label{hydro8}
\hbar\frac{\partial \tilde{\theta}_{+}}{\partial t}&=& -\hbar {v}_{+}^x \partial_x\tilde{\theta}_{+}-\hbar {v}_{+}^y \partial_y \tilde{\theta}_{+}-g_{2D}(\tilde{n}_{-}+\tilde{n}_{+})\label{hydro9}\\
\hbar\frac{\partial \tilde{\theta}_{-}}{\partial t}&=& -\hbar {v}_{-}^x\partial_x\tilde{\theta}_{-}-\hbar {v}_{-}^y \partial_y\tilde{\theta}_{-}-g_{2D}(\tilde{n}_{-}+\tilde{n}_{+}) \label{hydro10}
\end{eqnarray}
where ${v}_{\kappa}^x = \frac{\hbar}{m}\partial_x \bar{\theta}_\kappa- \kappa \eta ,
  \hspace{0.2cm}v_{\kappa }^y\hspace{0.1cm} =\hspace{0.1cm}\frac{\hbar}{m_y} \partial_y \bar{\theta}_\kappa$ are the background velocities of the components.

Rewriting the above set of equations in terms of total density $n_d=n_++n_-$, polarisation density $s_z=n_+-n_-$, total phase $\theta_d = \theta_+ - \theta_-$ and relative phase $\theta_r = \theta_+ - \theta_-$ and considering the limit $\bar{n}_{d} >> \bar{s}_{z}$, as the spin density for this case turn out to be relatively smaller than total density, Eqs(\ref{hydro7}-\ref{hydro10}) takes the following form:
\begin{widetext}
\begin{eqnarray}
\frac{\partial \tilde{n}_{d}}{\partial t}&=&-\partial_x  [\frac{\hbar \bar{n}_{d}}{2m}  \partial_x \tilde{\theta}_{d}+ \frac{\hbar \bar{s}_{z}}{2m}  \partial_x \tilde{\theta}_{r}]-\partial_x \bigg[ v^x  \tilde{n}_{d}+({v}_{s_{z}}^x-v^x) \frac{ \bar{s}_{z} }{\bar{n}_{d}}\tilde{s}_{z} \bigg]\nn\\ 
&-& \partial_y  [\frac{ \hbar \bar{n}_{d}}{2m_y}  \partial_y \tilde{\theta}_{d}+ \frac{\hbar \bar{s}_{z}}{2m_y}  \partial_y \tilde{\theta}_{r}]-\partial_y \bigg[v^y \tilde{n}_{d}+({v}_{s_{z}}^y-v^y)\frac{ \bar{s}_{z} }{\bar{n}_{d}}\tilde{s}_{z} \bigg]\label{hydro11}\\
\frac{\partial \tilde{s}_{z}}{\partial t}&=&-\partial_x  [\frac{\hbar \bar{s}_{z}}{2m}  \partial_x \tilde{\theta}_{d}+ \frac{\hbar \bar{n}_{d}}{2m}  \partial_x \tilde{\theta}_{r}]-\partial_x \bigg[ ({v}_{s_{z}}^x-v^x)\frac{ \bar{s}_{z} }{\bar{n}_{d}} \tilde{n}_{d}+  v^x s_{zp} \bigg]\nn\\ 
&-& \partial_y  [\frac{\hbar \bar{s}_{z}}{2m_y}  \partial_y \tilde{\theta}_{d}+ \frac{\hbar \bar{n}_{d}}{2m_y}  \partial_y \tilde{\theta}_{r}]-\partial_y \bigg[ ({v}_{s_{z}}^y-v^y) \frac{\bar{s}_{z}}{\bar{n}_{d}}  \tilde{n}_{d}+  v^y  \tilde{s}_{z} \bigg]\label{hydro12}\\
\hbar \frac{\partial \tilde{\theta}_{d}}{\partial t}&=& -\hbar \bigg[ v^x  \partial_x \tilde{\theta}_{d} 
 + v^y  \partial_y \tilde{\theta}_{d} 
-(v_{s_{z}}^x-v^x)\frac{ \bar{s}_{z} }{\bar{n}_{d}}  \partial_x \tilde{\theta}_{r} - (v_{s_{z}}^y-v^y)\frac{ \bar{s}_{z} }{\bar{n}_{d}} \partial_y \tilde{\theta}_{r}\bigg] -2 g_{2D} \tilde{n}_d\label{hydro13} \\ 
 \hbar\frac{\partial \tilde{\theta}_{r}}{\partial t} &=& -\hbar \bigg[ (v_{s_{z}}^x-v^x)  \frac{ \bar{s}_{z} }{\bar{n}_{d}} \partial_x \tilde{\theta}_{d} 
 + (v_{s_{z}}^y-v^y)\frac{ \bar{s}_{z} }{\bar{n}_{d}} \partial_y \tilde{\theta}_{d} 
 - v^x  \partial_x \tilde{\theta}_{r} - v^y \partial_y \tilde{\theta}_{r}\bigg] \label{hydro14}\\ \nn
\end{eqnarray}
\end{widetext}
where,
\bea
{\bs v}&=&\frac{n_+ {\bf v_{+ }} + n_- {\bf v_{- }}}{n_d}\nn, \hspace{0.1cm}
{\bf v}_{s_z}=\frac{n_+ {\bf v}_{+ } -  n_- {\bf v_{-}}}{s_z}\nn
\eea

are the velocities corresponding to total density and polarization density modes respectively. Neglecting the terms of the order of $\frac{ \bar{s}_{z} }{\bar{n}_{d}} $ times fluctuations, Eqs(\ref{hydro11}-\ref{hydro14}) can be written compactly as:
\begin{eqnarray}
\dot{\tilde{\rho}}&=&-\partial_x ( D_1 \partial_x\tilde\Theta) -\partial_y (D_2 \partial_y \tilde\Theta) + \nabla \cdot ( \bf{V} \tilde\rho)\label{rhoeq}\\
\dot{\tilde{\Theta}}&=&-{\bf{V}} \cdot \nabla \tilde\Theta - \frac{m {c^{x}}^2}{\hbar  \bar{n}_{d}}G \tilde\rho \label{theq}
\end{eqnarray}
Where,
\[ \tilde\rho  =
  \begin{bmatrix}
   \tilde{n}_{d} \\
    \tilde{s}_{z}
  \end{bmatrix}
 ,
 \tilde\Theta  =
  \begin{bmatrix}
    \tilde{\theta}_{d} \\
    \tilde{\theta}_{r}
  \end{bmatrix}
   , D_1 =
 \frac{\hbar}{2m } \begin{bmatrix}
  \bar{n}_{d} & \bar{s}_{z} \\
  \bar{s}_{z} &  \bar{n}_{d} \\
  \end{bmatrix}
   , \] \[  D_2 =
 \frac{\hbar}{2m_y } \begin{bmatrix}
  \bar{n}_{d} & \bar{s}_{z} \\
  \bar{s}_{z} &  \bar{n}_{d} \\
  \end{bmatrix}
   ,G =
 2 g_{2D} \begin{bmatrix}
 1&0\\
  0&0\\
  \end{bmatrix}
  \text{and}, \hspace{0.1cm} {\bf V}
  =  \begin{bmatrix}
   {\bs v} &  0 \\
   0 &   {\bs v} \\
  \end{bmatrix}.\]
  
Substituting $\tilde\rho=  \bigg(-\frac{\hbar \bar{n}_{d}}{m {c^x}^2 }\bigg)\bigg[    {G}^{-1}\dot{\tilde{\Theta}}+{G}^{-1} {\bf V} \cdot \nabla \tilde\Theta\bigg]$ from Eq.
  (\ref{theq}) in Eq.(\ref{rhoeq}) and simplifying, yields the following equation for phase fluctuations in the density modes $\tilde{\theta}_d$,
   \begin{equation}\label{oureq}
\partial_{\mu}(f^{\mu\nu}\partial_{\nu} \tilde{\theta_d})=0, \end{equation}
where
\[f^{\mu\nu}  = \frac{\bar{n}_{d}}{{c^x}^2 m}
  \begin{bmatrix}
   -1&  -v^x &  -v^y \\
  -v^x & {c^x}^2-{v^x}^2 &-v^x v^y \\
   -v^y & -v^y v^x &{c^y}^2-{v^y}^2\\
   \end{bmatrix},\]

   The scalar field satisfies the equation, $\frac{1}{\sqrt{-g}}\partial_{\mu}( \sqrt{-g} g^{\mu\nu}\partial_{\nu} \phi)=0$. Comparing it with Eq.(\ref{oureq}) gives, $\sqrt{-g} g^{\mu\nu} = f^{\mu\nu}$. Therefore, we get 
    \[g^{\mu\nu} = \frac{m m_y}{{\bar{n}_{d}}^2}
   \begin{bmatrix}
   -1&  -v^x &  -v^y \\
  -v^x & {c^x}^2-{v^x}^2 &-v^x v^y \\
   -v^y & -v^y v^x &{c^y}^2-{v^y}^2\\
   \end{bmatrix}\]
   with its inverse metric,
  \[g_{\mu\nu} = \bigg(\frac{{\bar{n}_{d}}^2}{m m_y {c^x}^2}\bigg) 
  \begin{bmatrix}
  - ({c^x}^2 -{v^x}^2-\frac{{v^y}^2}{\alpha})&  -{v^x} & \frac{-{v^y}}{\alpha} \\
  -{v^x} &  1 &0\\
 \frac{-{v^y}}{\alpha} & 0 &\frac{1}{\alpha}\\
   \end{bmatrix}\]
   
   where $\frac{{c^x}^2}{{c^y}^2}=\frac{1}{\alpha}$ and $det f^{\mu\nu}=\sqrt{-g}$.
 Thus, the line element can be written as:   
\begin{eqnarray}
ds^2&=&g_{\mu\nu} dx^{\mu}dx^{\nu}\nonumber\\
 &=&\bigg(\frac{{\bar{n}_{d}}^2}{m m_y {c^x}^2}\bigg) \bigg\{-\bigg({c^x}^2 -\bigg[ (\frac{v^y}{\sqrt{\alpha}})^2+{v^x}^2\bigg]\bigg) dt^2\nn\\&-&2 \bigg[{v^x} dx +\frac{v^y}{\sqrt{\alpha}}\frac{dy}{\sqrt{\alpha}}  \bigg] dt + (dx^2 + (\frac{dy}{\sqrt{\alpha}})^2 )\bigg\},\nonumber
\end{eqnarray}
which in terms of total sound velocity, $c_s$ yields Eq.(\ref{metric}) of the main text.

\begin{figure}[htbp]
\includegraphics[scale =0.75]{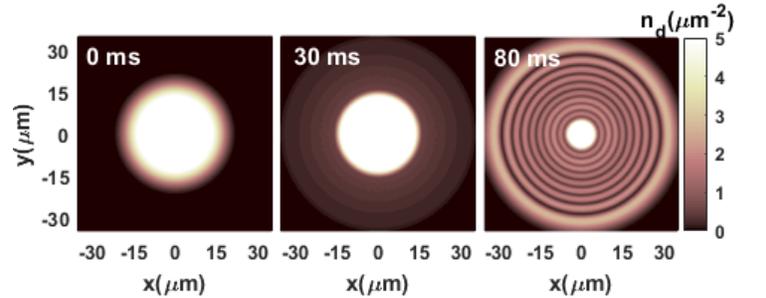}
\caption{{\it (color online).} Time evolved density for a two-component BEC.}\label{scalar_te}
\end{figure}
\begin{figure}[htbp]
\includegraphics[scale =0.68]{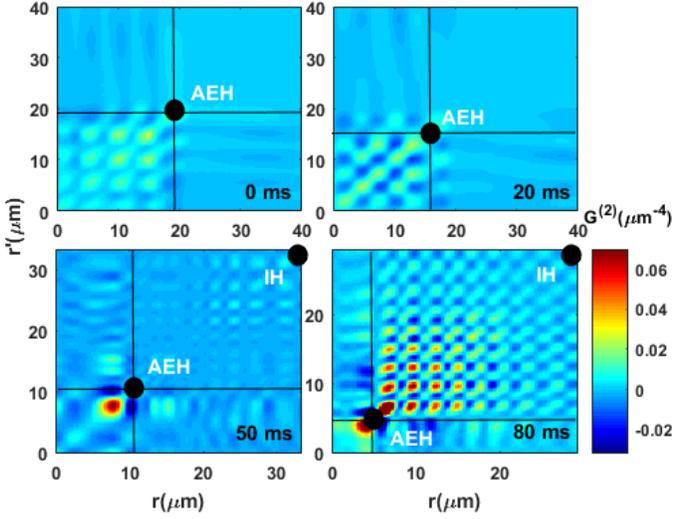}
\caption{{\it (color online).} Density-density correlation for a two-component BEC $G^{(2)}(r,r';0^\circ)$ without SOC, obtained using TWA, is shown at various times.}\label{tc_corr}
\end{figure}
\begin{figure*}
\includegraphics[width=14cm,height=7cm]{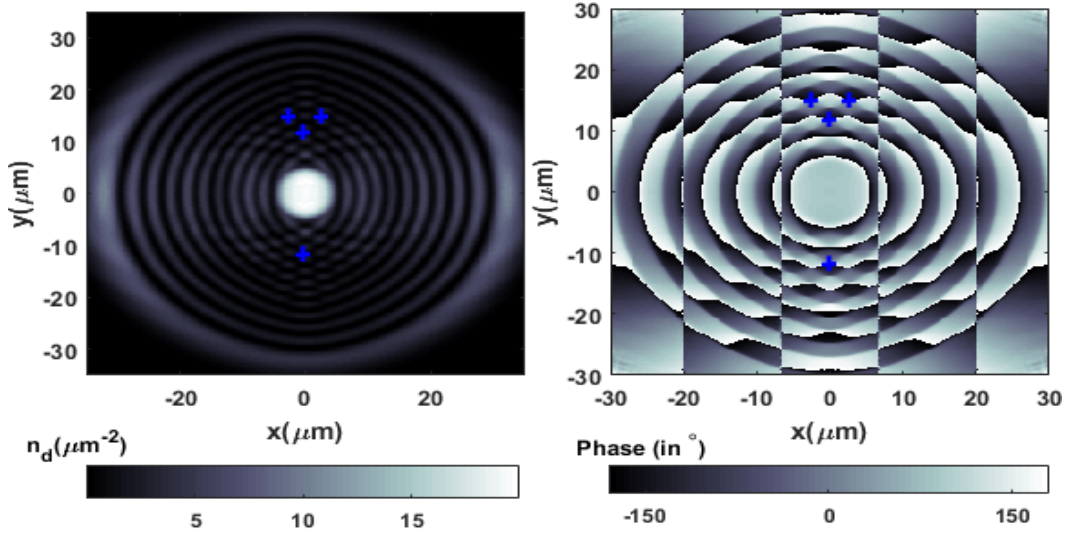}
\caption{{\it (color online).} Total density (left) and its corresponding phase variation (right) for $\eta'/\eta=0.78$ at 80 ms. The marker `+' in the phase shows 
the presence of vortices at few of the marked minima location of the left figure, where the interference fringes are formed.}\label{phase}
\end{figure*}
In order to compare the results presented in this work with a non-SOC case, we take a two-component BEC \cite{pitaevski} with equal interaction strengths (= $g_{2D}$).
The GPE to be solved for this case in the potential profile, given in Eq.(\ref{pot}), is:\\
\bea 
 i\hbar\frac{\partial \psi_{\kappa}  }{\partial t} &=& \bigg[-\frac{\hbar^2}{2m}\partial_x^2-\frac{\hbar^2}{2m}\partial_y^2 + V_{2D}(\bs{r},t)\bigg]\psi_{\kappa}\nn\\
 & +& g_{2D}( |\psi_+|^2+|\psi_-|^2)\psi_{\kappa},~ \kappa=\pm     \label{gpe_tc} 
\eea

The time evolved density for a two-component BEC, i.e. the case without any spin-orbit coupling in the system is represented in Fig.\ref{scalar_te}. The components of flow velocity corresponding to the density mode, in this case, are given as,
\bea
{v_d}^x &=& \frac{{\hbar}}{2 i m n_d}\bigg( \bs{\Psi^\dagger} \partial_x \bs{\Psi} - \bs{\Psi^T} \partial_x \bs{\Psi^*} \bigg)\nn\\
{v_d}^y &=& \frac{{\hbar}}{2 i m n_d}\bigg(\bs{\Psi^\dagger} \partial_y \bs{\Psi} - \bs{\Psi^T} \partial_y \bs{\Psi^*} \bigg)\label{vel2c}
\eea

\section{Thermal Spectra of Hawking radiation}\label{TSHR}

\begin{figure}
\includegraphics[scale=0.80]{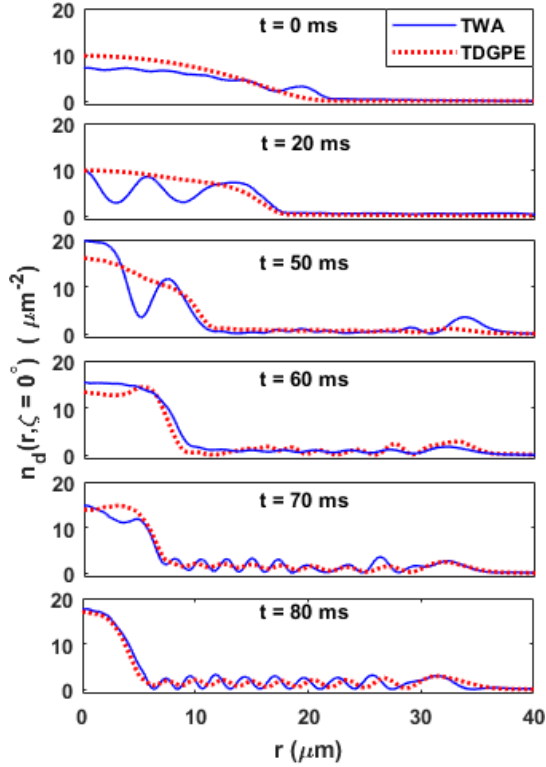}\\
\caption{{\it (color online).} A cross-sectional profile of time-dependent GPE (TDGPE) density evolution superimposed with those obtained from TWA along $\zeta = 0^\circ$ using Fig.\ref{MF_GPE}(a) and Fig.\ref{TWA_GPE}. In both the cases, amplification of density in the supersonic region (where density modulations appear), is observed. However, in this region, slight shifts in the amplitude were observed in the density profiles obtained using TWA.}\label{1dtwa_gpe}
\end{figure}

 In this appendix, we will provide the details of the formalism developed for in-situ measurements in the trapped condensate \cite{entanglementJS,stein_phonon}, to extract the spectrum information from the evaluation of density-density correlation function. The theory was developed for a single component BEC in ref. \cite{entanglementJS} and we, therefore, generalise it for our system and use the corresponding equations for the density modes. The following method can be employed for trapped condensate, discussed in this work, but only in certain regions where we have approximately homogenous upstream and downstream region. The region is chosen near the Hawking pair correlation line in the density-density correlation, shown in Fig.\ref{thermala}(a,b). We will use the density -density correlation function obtained using the TWA method. A comparison of the cross-sectional profile of the total density obtained using TDGPE and from TWA method (Fig.\ref{MF_GPE}(a) and Fig.\ref{TWA_GPE}) is shown in Fig. \ref{1dtwa_gpe}. 
 
   Corresponding to the density modes, the Fourier transform of the density operator,\bea
\rho_k = \sum_p \hat{a}^\dagger_{p+k}\hat{a}_{p},\label{FTD}
\eea
where $\hat{a}_{p}$ is the annihilation operator for a single atom with momentum $\hbar$p. 

The annhilation operator corresponding to the density modes : $\hat{a}_{k} = u_{k} \hat{b}_{k}+ v_{k} \hat{b}^\dagger_{-k}$,
where
\bea
u_{k} &=& u_{1k} + u_{2k} = \sqrt{\frac{\xi^k}{{\varepsilon_d}^k}+1},\nn\\
v_{k} &=& v_{1k} + v_{2k} = \sqrt{\frac{\xi^k}{{\varepsilon_d}^k}-1}\nn
\eea
with,

 $\xi^k = E^{2D}_-({\bf k}) + g{n}_d,
 \varepsilon_d = \sqrt{E_{-}^{2D}(\bs{k})[E_{-}^{2D}(\bs{k}) +2 g {n}_d]}$
 and,
 $E^{2D}_-({\bf k}) = \frac{\hbar^2k_x^2}{2m} + \frac{\hbar^2k_y^2}{2m_y}$.

The eq.(\ref{FTD}) in Bogoliubov approximation ($\hat{a}_0, \hat{a}^\dagger_0 \rightarrow \sqrt{N}$) can be written as: $
\rho_k = \sqrt{N} (u_k + v_k) (\hat{b}^\dagger_k + \hat{b}_{-k}),$
N is the total number of atoms and $u_k$, $v_k$ are the Bogoliubov amplitudes. The densities in the upstream and the downstream regions can be written as:
$\rho_k^u = \sqrt{N^u}( u_{k_{H}} + v_{k_{H}})({{\hat{b}^{u^\dagger}}_{k_{H}}} + {\hat{b}^u}_{-k_{H}} ),
\rho_k^d = \sqrt{N^d}( u_{k_{P}} + v_{k_{P}})({{\hat{b}^{u^\dagger}}_{k_P}} + {\hat{b}^u}_{-k_P} )
$. Thus, we get:
\bea
&<\rho^i_{k_i}\rho^j_{k_j}>& = \sqrt{N^i N^j} ( u_{k_{i}} + v_{k_{i}})( u_{k_{j}} + v_{k_{j}})[ <{{\hat{b}^{i^\dagger}}_{k_{i}}}{{\hat{b}^{j^\dagger}}_{k_{j}}} > + \nn\\
&<{{\hat{b}^{i^\dagger}}_{k_{i}}}{{\hat{b}^{j}}_{-k_{j}}} > &+ <{{\hat{b}^{i^\dagger}}_{-k_{i}}}{{\hat{b}^{j}}_{k_{j}}} >  + <{{\hat{b}^{i^\dagger}}_{-k_{i}}}{{\hat{b}^{j}}_{-k_{j}}} > + \delta_{ij} \delta_{-k_{i}k_{j}}]\nn\\\label{rho1rho2}
\eea
where i,j = u/d. 
Considering a specific case for extracting the information about spectrum from the correlations between Hawking and its partner modes in (u,d) region, i.e taking  i = u, j = d, $k_i = - k_{H}$ and, $k_j = - k_{P}$ in the Eq.(\ref{rho1rho2}): 
\bea
&<\rho^u_{-k_{H}}\rho^d_{-k_{P}}>& = \nn\\
&\sqrt{N^u N^d}& ( u_{k_{H}} + v_{k_{H}})( u_{k_{P}} + v_{k_{P}})[ <{{\hat{b}^{u^\dagger}}_{-k_{H}}}{{\hat{b}^{d^\dagger}}_{-k_{P}}} >  + \nn\\
&<{{\hat{b}^{u^\dagger}}_{-k_{H}}}{{\hat{b}^{d}}_{k_{P}}} > & + <{{\hat{b}^{u^\dagger}}_{k_{H}}}{{\hat{b}^{d}}_{-k_{P}}} > + <{{\hat{b}^{u}}_{k_{H}}}{{\hat{b}^{d}}_{k_{P}}} > ]\label{rho1rho3}
\eea
Assuming the number of excitations travelling with the flow (with negative k) is very small and are negligible. Thus, setting any term with negative ``k" value to zero, we get:
\bea
&<\rho^u_{-k_{H}}&\rho^d_{-k_{P}}> = \nn\\ 
&\sqrt{N^u N^d}& ( u_{k_{H}} + v_{k_{H}})( u_{k_{P}} + v_{k_{P}}) <{{\hat{b}^{u}}_{k_{H}}}{{\hat{b}^{d}}_{k_{P}}} > \label{rho1rho2f}\nn\\
\eea


Since, here we have calculated the spectral information for the spectrum along  $\zeta = 0^{\circ} $ (i.e along the x-direction). Also, in order to make use of the density-density correlation data, we use the Fourier transform definition of density, $
\rho^i_{k_i} = \int dr e^{-ik_i  r}  n^i(r)$.
Here, we are evaluating the spectral information only for the `k' values that are in the direction of r only i.e in radial direction only i.e ${\bf k_i} \cdot {\bf r} = k_i \hat{r} \cdot \hat{r} r = k_i r$.   
Therefore,
$$<\rho^i_{k_i}\rho^j_{k_j}> = \int dr dr' e^{-ik_i  r} e^{-ik_j  r'} <n^i(r)n^j(r')>$$. 

Thus, $<\rho^i_{k_i}\rho^j_{k_j}>$ can be measured by computing the Fourier transform of the density-density correlation function. Putting $k_i = -k_{H}, k_j = -k_P$ results in the following:
\bea
<\rho^u_{-k_{H}}\rho^d_{-k_P}> = \int dr dr' e^{ik_{H}  r} e^{ik_P  r'} <n^u(r)n^d(r')>\label{rho1rho2_2}\nn\\\eea
Comparing the above expression with Eq.(\ref{rho1rho2f}), we obtain Eq.(\ref{radspec}) of section \ref{SOR}. 

In the last part of this appendix, we will provide the details of evaluation of Hawking temperature $T_{H}$ at the acoustic event horizon of this analogue black hole using the gravitational analogy, by generalizing the definition of $T_{H}$  used for SBH in a scalar condensate. Its determination depends on surface gravity and is given as,
$T_H= \frac{\hbar g_H}{2 \pi k_B c_H}$ \cite{visser11,Steinhauer14, Steinhauer18}. In general,
the "surface gravity" is given as:
\begin{equation}
g_H=\frac{1}{2}\frac{d(c_H^2-{v_{\perp}} ^2)}{dn}\bigg|_{hz}
\end{equation}
where $c_H = \frac{{c_s}}{\sqrt{(1+\alpha)}}$ and $v_{\perp}= {\bf v} \cdot \hat{n}$ is the normal component of velocity \cite{visser90}. Thus, the Hawking temperature becomes, 
\bea
T_H=\frac{\hbar }{2 \pi k_B}\bigg( \frac{d}{dn}[\frac{{c_s}}{\sqrt{(1+\alpha)}}-{{\bf v'}\cdot \hat{n}}]+ \frac{d}{dn}[{\bf g}(\eta,\eta')\cdot \hat{n}]\bigg)\bigg|_{hz} \label{HT}
\eea
where $`\it{n}$' corresponds to the spatial coordinate normal to the horizon ($\it{hz}$); $ {\bf v}'$ and ${\bf g}(\eta,\eta')$ have the form as follows:
\bea {\bf v}' &=& \frac{{\hbar}}{2 i m n_d}\bigg( \bs{\Psi^\dagger} {\bf \nabla} \bs{\Psi} 
+ \bs{\Psi^T} {\bf \nabla}  \bs{\Psi^*} \bigg) \nn\\
{\bf g}(\eta,\eta')&=& \hat{x} \frac{\eta}{n_d} \bs{\Psi^\dagger}  \check{\sigma}_z \bs{\Psi} +\hat{y}\frac{{\hbar {\eta'}^2}}{2 i m n_d{\eta}^2}\bigg( \bs{\Psi^\dagger} \partial_y \bs{\Psi} + \bs{\Psi^T} \partial_y \bs{\Psi^*} \bigg)\nn
\eea

The Eq. (\ref{HT}), explicitly shows the contribution of the SOC on analogue Hawking temperature. Furthermore, $g_H$ and the Hawking temperature $T_H$, in the SBH formed out of SOC-BEC exhibits direction-dependent behaviour due to the anisotropic nature of sound and flow velocity [Eq. (\ref{velocities})] in this configuration, as evident from the second term of Eq. (\ref{HT}). The purpose of the calculation of analogue Hawking temperature, in section \ref{SOR}, was to check the accuracy of the results obtained using the correlation function in Eq.(\ref{radspec}) with that obtained from the usage of the definition of $|\beta|^2$ relying on $T_H$. The evaluation is only done at the initial times where the sonic analogy is valid and thus, show that the computed spectrum is thermal. The formula of $T_H$ in Eq.(\ref{HT}) is strictly valid in the hydrodynamic regime where the sonic analogy, is preserved. Thus, we have used it to evaluate the predicted thermal spectrum only at the initial times in the simulation.

   \end{document}